\documentclass[11pt,twocolumn]{article}
\usepackage[]{amsmath}
\usepackage[ruled]{algorithm2e}
\usepackage[switch]{lineno}
\usepackage{geometry}
\usepackage{graphicx}  
\usepackage{subfigure}  
\usepackage{epstopdf}
\usepackage{microtype}
\usepackage[ngerman,english]{babel}
\usepackage{caption}
\usepackage[utf8]{inputenc}
\usepackage{sectsty}
\usepackage{color}
\usepackage{relsize}
\usepackage{appendix}
\usepackage{siunitx}

\allsectionsfont{\normalfont\sffamily\bfseries}
\usepackage{amssymb, trfsigns,multicol,upgreek,nicefrac,dsfont,bm,cancel,amsbsy}
\usepackage{mathrsfs}
\usepackage{url}

\usepackage[colorinlistoftodos,prependcaption,textsize=small]{todonotes}
\usepackage{color}
\usepackage{colortbl}
\usepackage{transparent}
\usepackage{hyperref}
\hypersetup{
	colorlinks,
	linkcolor={blue!80!black},
	citecolor={blue!80!black},
	urlcolor={blue!80!black}
}
\addto\extrasenglish{

}

\usepackage{fancyhdr}
\usepackage{lipsum}

\makeatletter
\def\@maketitle{%
	\newpage
	\null
	\begin{center}%
		\let \footnote \thanks
		{\LARGE \@title \par}%
		\vskip 1.5em%
		{\large
			\lineskip .5em%
			\begin{tabular}[t]{c}%
				\@author
			\end{tabular}\par}%
		\vskip 1em%
		{\large \@date}%
	\end{center}%
	\par
	\vskip 1.5em}
\makeatother

\makeatletter
\def\@maketitle{%
	\newpage
	\null
	\vspace*{-1cm}
	\begin{center}%
		\let \footnote \thanks
		{\LARGE \@title \par}%
		\vskip 1.5em%
		{\large
			\lineskip .5em%
			\begin{tabular}[t]{c}%
				\@author
			\end{tabular}\par}%
		\vskip 1em%
		{\large \@date}%
	\end{center}%
	\par
	\vskip 0.5em}
\makeatother

\title{Diffuse Sound Field Synthesis
}
\author{\normalsize Franz Zotter, Stefan Riedel, Lukas Gölles, Matthias Frank\\ \normalsize Institute of Electronic Music and Acoustics\\ \normalsize University of Music and Performing Arts Graz, Austria}
\date{}

\begin{document}
\selectlanguage{english}
 \nolinenumbers
\twocolumn[
\begin{@twocolumnfalse}
	\maketitle
\begin{abstract}\noindent
    Can uncorrelated surrounding sound sources be used to generate extended diffuse sound fields? By definition, targets are a constant sound pressure level, a vanishing average sound intensity, uncorrelated sound waves arriving isotropically from all directions. Does this require specific sources and geometries for surrounding 2D and 3D source layouts?   \\ 
    As methods, we employ numeric simulations and undertake a series of calculations with uncorrelated circular/spherical source layouts, or such with infinite excess dimensions, and we point out relations to potential theory. Using a radial decay $\nicefrac{1}{r^\beta}$ modified by the exponent $\beta$, the representation of the resulting fields with hypergeometric functions, Gegenbauer polynomials, and circular as well as spherical harmonics yields fruitful insights.
    \\
    In circular layouts, waves decaying by the exponent $\beta=\nicefrac{1}{2}$ synthesize ideally extended, diffuse sound fields; spherical layouts do so with $\beta=1$. None of the layouts synthesizes a perfectly constant expected sound pressure level but its flatness is acceptable. 
    Spherical $t$-designs describe optimal source layouts with well-described area of high diffuseness, and non-spherical, convex layouts can be improved by restoring isotropy or by mode matching for a maximally diffuse synthesis.
    Theory and simulation offer a basis for loudspeaker-based synthesis of diffuse sound fields and contribute physical reasons to recent psychoacoustic findings in spatial audio.
\end{abstract}
\vspace*{0.5cm}
\end{@twocolumnfalse}
]

\section{Introduction} 
The concept of a diffuse sound field refers to uncorrelated unit-variance plane waves impinging from all directions. In an `ideal', cylindrically or spherically isotropic diffuse field, the sound pressure level is position-independent and the time-averaged sound intensity vector vanishes at all positions \cite{jacobsen}.
Measurement of sound field diffuseness has an extensive history in acoustics, for instance based on the well-defined frequency-dependent correlation between two points in space \cite{cook1955measurement, balachandran1959random, kuttruff1963raumakustische, dammig1957messung}. In addition, the behaviour of the sound intensity vector can be used to quantify the diffuseness of the sound field \cite{veit1987production, pulkki2007spatial}. 
In particular, sound intensity normalized by the energy density and speed of sound were defined as energy vector~\cite{Gerzon92,Merimaa2007} when the diffuse sound field is synthesized by incoherently driven loudspeakers, as it  simplifies to the gain-square-weighted average of the loudspeaker directions.  The resulting vector is of unit length for fully directional sound fields and zero length for fully diffuse sound fields. 
More recent approaches employed spherical microphone arrays to measure the diffuseness of the sound field, e.g.\ COMEDIE, which is based on an eigenvalue decomposition of the spherical harmonics (SH) covariance matrix \cite{epain2016spherical}.
A directional and temporal histogram of the energy decay in a simulated room was proposed in auralization~\cite{Schroeder,vorlaender}. 
Several recent works moreover showed ways to retrieve or represent anisotropic reverberation properties based on measurements~\cite{merimaa2005sirr,tervo,nolan,berzborn,Masse,Alary,romblom,coleman,Goetz,Hold,deppisch}.

Synthesis of diffuse sound fields with loudspeaker arrays is of interest in laboratory environments, for example to quantify the diffuse-field sound transmission of building partitions~\cite{hoorickx2022} or acoustic absorption~\cite{robin2014,Dupont23}, the diffuse-field response of microphone arrays~\cite{habets,kustner,akbar2020} and dummy heads \cite{theile1984comparison, mckenzie2018diffuse, armstrong2018perceptual, NeumannKU100Manual}, or to measure the performance and perceptual quality of active noise control/cancellation algorithms \cite{moreau2009active, holzmuller2023frequency, elliott1988active}, etc.\

Spatial sound reproduction targets a faithful reproduction of diffuse sound fields that perceptually  elicits a sensation of envelopment for the audience, by suitably employing surrounding loudspeakers \cite{walther2011assessing, Hiyama2002, cousins2015subjective, cousins, riedel2023effect}.

Theoretically, sound field synthesis techniques such as wave field synthesis (WFS) \cite{berkhout,start,verheijen,caulkins,ahrens,firtha_referencing,winter,grandjean1} or near-field compensated higher-order Ambisonics (NFC-HOA) ~\cite{nicol,daniel,ward,poletti,ahrens,grandjean1} should allow to reproduce an ideally diffuse sound field, given a continuous distribution of elementary sources.
In discrete-direction implementations, however, spatial aliasing severely affect the sound field synthesis capacities of both WFS and HOA in much of the auditory range. Nevertheless, it could be argued that the fruitfulness of these approaches lies in much more tolerant psychoacoustic effects of  auditory localization~\cite{spors2013}, so that playback qualities allow for a much larger sweet area, as analyzed in experiments and perceptual measures \cite{frank,wierstorf,stitt,kuntz1,kuntz2}.

In WFS, it was studied for diffuse-field synthesis how many uncorrelated plane-wave sources are needed for a centered listener~\cite{sonke,ahrensDiffuse}. In Ambisonics, the required directional resolution or Ambisonic order for a large sweet area was investigated~\cite{frank2017},
and a directionally non-uniform reverberation algorithm was suggested~\cite{Alary2019}. 

The minimum number of loudspeakers and horizontal arrangements for reproducing the spatial impression of diffuse sound fields in channel-based playback was investigated in terms of interaural cross correlation \cite{Damaske1972}  and listening experiments \cite{Hiyama2002} with uncorrelated signals.

Interaural coherence and interaural level differences were moreover considered to evaluate the quality of diffuse-field reproduction capabilities~\cite{walther2011assessing,cousins}.\\

Adding height loudspeaker layers slightly increases perceived diffuseness, when diffuseness is defined as `\emph{the sound coming from all directions with equal intensity. Therefore, the sound should ideally be impossible to localise and without any gaps [...] in three dimensions}' as shown in \cite{cousins2015subjective, cousins2016}. Other researchers showed that two distinct spatial attributes related to the perception of diffuse sound fields can be separated, namely `\emph{envelopment}' for being surrounded by sound and `\emph{engulfment}' for being covered by sound \cite{riedel2023effect, sazdov2007perceptual}. 
\emph{Isotropy} as technical measure of directional uniformity was recently discussed~\cite{nolan}, and spherical $t$-designs were shown to be optimal plane-wave layouts to synthesize isotropic sound fields~\cite{tanaka23}\\
Reverberant spatial audio objects and diffuse-field modeling were recently proposed \cite{romblom,coleman} to model the typical/anisotropic characteristics of acoustic rooms.
Bandpass-based decorrelation for Diffuse Field Modeling that improves diffuseness in measured $1^\mathrm{st}$-order Ambisonic room impulse responses was proposed. By manipulation of the directional weights for front, left, and right, the perception of envelopment was shown to be affected to some degree by direction dependent diffuse fields~\cite{romblom}. Further investigating the perceivability of anisotropy in reverberation, 
i.e., by recognizing a rotation of the scene, revealed that an time average of the interaural energy decay difference at 850~Hz of slightly more than $\SI{1}{dB}$ can already be heard~\cite{Alary}.
Simplified auralization of non-uniform room reverberation was moreover proposed and tested~\cite{Kirsch}, using a warped, sparse layout of virtual reverberation sources.\\

In loudspeaker-based reproduction~\cite{bs.2051}, channel-based, scene-based (Ambisonics), and object-based amplitude panning concepts mostly employ point-source loudspeakers for playback, yielding a $1/r$ sound pressure decay with distance and parallax with regard to the listening position. Object-based WFS loudspeaker systems are typically 2.5D and based on dense horizontal point-source loudspeaker arrays. When rendering plane-wave sound objects, they yield a $1/\sqrt{r}$ sound pressure decay with distance, depending on  the listening position, cf.~\cite{ahrens,grandjean1}. 
Surprisingly, all typical loudspeaker-based reproduction still exhibits pronounced shortcomings when used to synthesize extended perceptually diffuse, and therefore consistently non-directional, sound fields:\\

Object-based rendering using amplitude panning, channel-based rendering, or scene-based Ambisonic rendering on moderateley dense loudspeaker layouts can be interpreted to exhibit a notoriously small sweet area regarding their capacity of rendering uniformly diffuse listener envelopment.
Recent studies revealed that instead of the conventional point-source loudspeakers, vertical line-source loudspeakers could supply a larger sweet area with a perceptually diffuse sound field~\cite{heidegger2020,blochberger2019sweet,riedel2022surrounding,Riedel2023}. With conventional point-source loudspeakers, the directional impression would often be dominated by the loudspeakers closest to the off-center listener location, even if the off-center displacement is limited to, say, a third of the layout radius.

A lot of research drive behind WFS was fueled by the idea that its many loudspeakers and well-controlled time delays avoid a confined sweet area by shaping consistent and extended wave fronts~\cite{spors2013}. Plane waves were thought of as an ideal virtual source type to consistently supply a large listening area. And yet, experiments using 2.5D WFS showed a clear benefit in diffuse-field rendering when the 2.5D reference line correcting the synthesized plane-wave levels was specific to a certain listening position~\cite{Melchior2008}. So also centered 2.5D wave-field synthesis appears limited in rendering a perceptually diffuse sound field to off-center listeners. 
WFS theory suggests~\cite{start,verheijen,ahrens,winter,firtha} that vertical line sources loudspeakers are the optimal sound sources for 2D WFS, of which the generated sound field is constant along the vertical dimension. As a consequence, 2D WFS does not require any correction and therefore no reference line, in contrast to 2.5D WFS. Nevertheless, the associated effort in light of the already horizontally dense loudspeaker arrangements appears enormous for 2D WFS compared to 2.5D WFS; this effort might only be realistic with local WFS systems~\cite{winter} that consider a reduced number of loudspeakers while accepting a limited audience area. \\

\begin{figure*}[t]
	\centering
   \subfigure[circle of point sources]{
 \includegraphics[height=37mm,trim=8mm 7mm 6mm 7mm,clip]{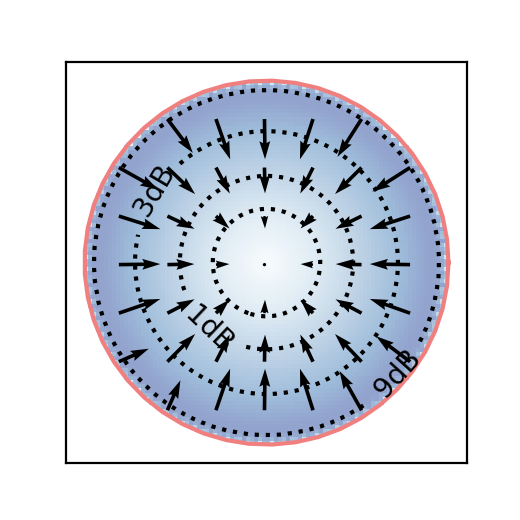}}
	\subfigure[circle of vert. line sources]{
 \includegraphics[height=37mm,trim=8mm 7mm 6mm 7mm,clip]{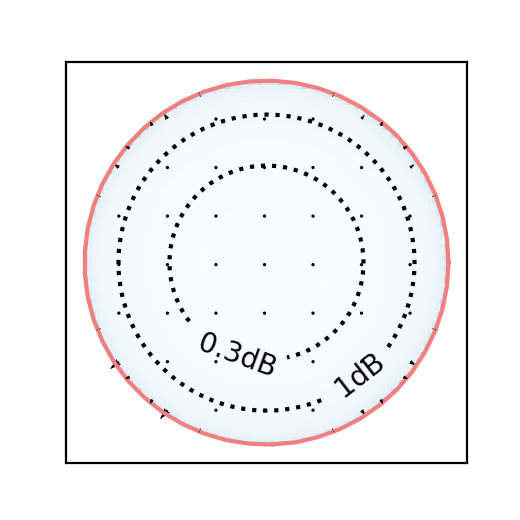}}
	\subfigure[circle of horiz. line sources]{
 \includegraphics[height=37mm,trim=8mm 7mm 6mm 7mm,clip]{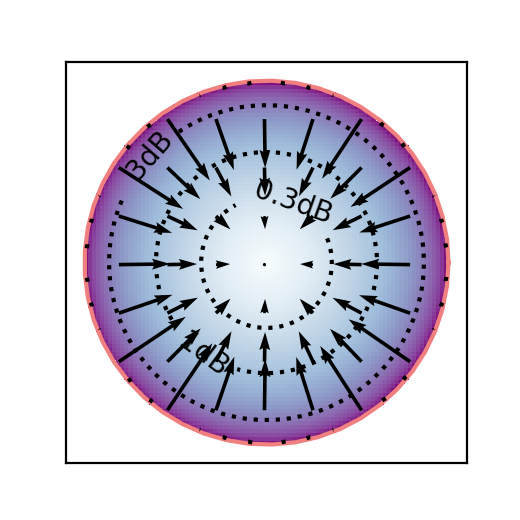}\quad}
 \hfill
	\subfigure[sphere of point sources]{\includegraphics[height=37mm,trim=8mm 7mm 6mm 7mm,clip]{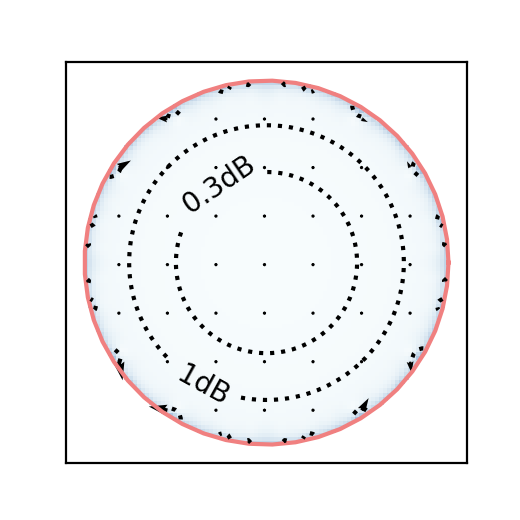}}
 \quad\includegraphics[height=38mm,trim=0mm 5mm 0mm 5mm,clip]{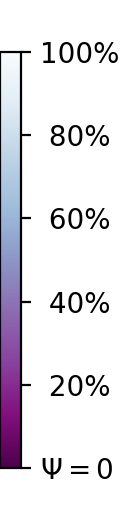}
	\caption{Horizontal map of sound energy density (contours), normalized intensity $\bm{I}/(c\,w)$ (arrows), and diffuseness $\psi$ (colors)
	with 100 regularly spaced sources on the azimuth for a circle (salmon), or for a sphere with 480 sources at Chebyshev-type nodes from Gr\"{a}f's website~\cite{graefurl}, all playing statistically independent signals of uniform variance.
	\label{fig:simu1}}
\end{figure*}

If vertical line-source loudspeakers also enlarge the diffuse-field rendering capacities in WFS, they appear to be similarly beneficial regardless of the particular kind of 2D audio rendering method. However, considering that only WFS synthesizes consistent parallel first wave fronts of large extent~\cite{spors2013,ahrens}, while wave fronts will rather be globally concentric otherwise, a common conclusion might be coincidental. Suitable wave fields produced by the playback loudspeakers therefore deserve a more fundamental investigation. Preferably this yields an explanation technical thorough enough to serve as an alternative to psychoacoustic experimentation or to psychoacoustic modeling based on interaural level difference (ILD) and interaural coherence (IC)~\cite{Riedel2023}. In this way it could also permit utilization, e.g., in technical acoustic applications.\\

A systematic theory is desirable, and might contribute to a profoundly better handling and understanding of diffuse sound field synthesis, which motivates our paper.

\subsection{Background and motivation}\label{sec:motivation}
A noteworthy quote by Jacobsen and Roisin describes diffuse sound fields   \cite{jacobsen2000coherence}:
\begin{quote}
    [\dots M]ost acousticians would agree on a definition that involves sound coming from all directions. This leads to the concept of a sound field in an unbounded medium generated by distant, uncorrelated sources of random noise evenly distributed over all directions. Since the sources are uncorrelated there would be no interference phenomena in such a sound field, and the field would therefore be completely homogeneous and isotropic. For example, the sound pressure level would be the same at all positions, and temporal correlation functions between linear quantities measured at two points would depend only on the distance between the two points. The time-averaged sound intensity would be zero at all positions. An approximation to this ‘‘perfectly diffuse sound field’’ might be generated by a number of loudspeakers driven with uncorrelated noise in a large anechoic room [\dots]
\end{quote}
As metrics~\cite{Kuttruff}, we employ sound energy density $w$, sound intensity $\bm I$, and diffuseness $\psi$ in the sound field~\cite{merimaa2005sirr,pulkki2007spatial}, which are described by the sound pressure $p$ and sound particle velocity vector $\bm v$, according to
\begin{align}
	 w&=\frac{E\{|p|^2\}}{2\rho\,c^2}+\frac{\rho E\{\|\bm v\|^2\}}{2}, &
 {\bm I}&=\frac{E\{\Re\{p^*\bm v\}\}}{\rho\,c},\\
 \psi&=1-\frac{\|\bm I\|}{c\,w},\label{eq:wIPsi}
\end{align}
with the density of air $\rho=\SI{1.2}{\frac{kg}{m^3}}$, the speed of sound $c=\SI{343}{\frac{m}{s}}$, and the statistical expectation $E\{\cdot\}$.
In a perfectly diffuse sound field, these quantities are expected to ideally assume, 
\begin{align}
	{w}&=\text{const},& {\bm I}&=\bm 0, &\psi&=1.
\end{align} 
Otherwise, diffuseness becomes $\psi=0$ for a non-diffuse sound field from a single source.

To observe whether $w$ as a measure of sound pressure level is constant and whether the average sound intensity $\bm I$ vanishes, ideally, we present a small simulation study motivating the particular research questions of this paper.

\autoref{fig:simu1} shows a simulated free-field map of the sound intensity (vectors), the diffuseness (colors), and the sound energy density (contours) along the horizontal plane, for circular or spherical surfaces $S$ of acoustic sources radiating uncorrelated signals of equal variance from many uniformly distributed directions $-\bm u$. 
The simulation uses simplified integrals \cite{Merimaa2007}, cf.~\autoref{sec:expectation},
\begin{align}
    \rho\,c^2 w&=\frac{1}{\hat N}\sum_{l=1}^\mathrm{L}\sigma_l^2\,|G_l|^2\,&
    \rho\,c\,\bm I&=\frac{1}{\hat N}\sum_{l=1}^\mathrm{L}\bm u_l\,\sigma_l^2\,|G_l|^2\label{eq:numeric}
\end{align}
determining $\psi$ in eq.~\eqref{eq:wIPsi}. 
The Green's function decays with $|G_l|\propto\nicefrac{1}{\sqrt{r_l}}$ for a line source and $|G_l|\propto\nicefrac{1}{r_l}$ for a point source,  cf.~eq.~\eqref{eq:green_hf}. Distance $r_l$ and normalized direction vector $\bm u_l$ are measured from each point or line source to the receiver; $\rho$ is the air density, $c$ the speed of sound, and $\hat N$ normalizes $\rho\,c^2\,w(\bm 0)=1$.
With the levels $\sigma_l=1$, vertical line sources arranged in a horizontal circle yield the mapping in \autoref{fig:simu1}~(b), tangentially arranged they yield (c), and a circle of point sources yields (a).
Point sources on a sphere yield the mapping (d).

\begin{figure*}[t]
	\centering
    	\subfigure[$\sigma=1$]{
 \includegraphics[height=37mm,trim=4mm 4mm 3mm 3mm,clip]{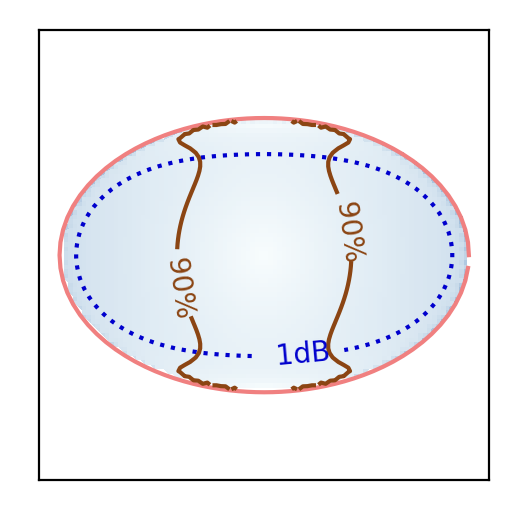}}	
 \subfigure[$\sigma=\sqrt{R}$]{
 \includegraphics[height=37mm,trim=4mm 4mm 3mm 3mm,clip]{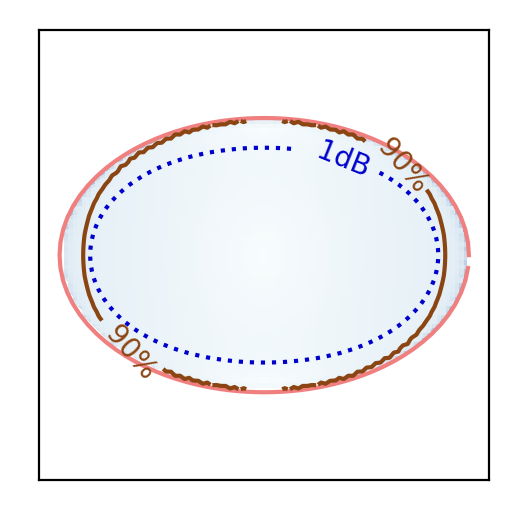}}
    	\hfill\subfigure[$\sigma=1$]{
 \includegraphics[height=37mm,trim=4mm 4mm 3mm 3mm,clip]{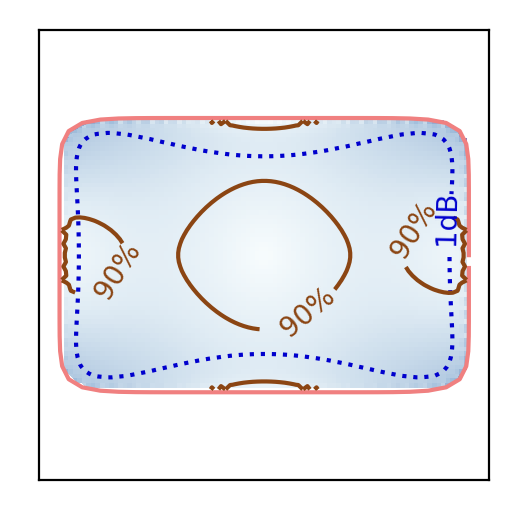}}	
 \subfigure[$\sigma=\sqrt{R}$]{
 \includegraphics[height=37mm,trim=4mm 4mm 3mm 3mm,clip]{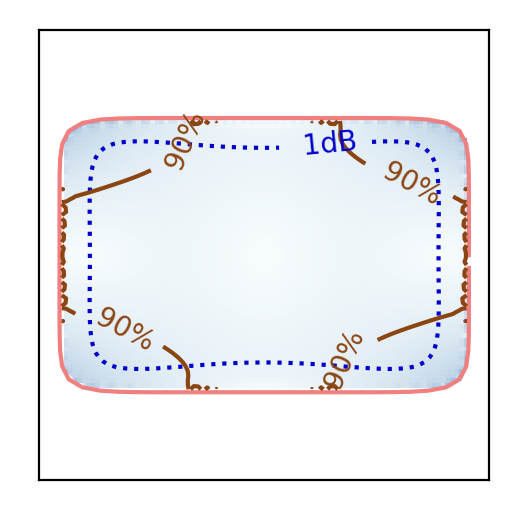}}
 \quad\includegraphics[height=38mm,trim=0mm 5mm 0mm 5mm,clip]{colorbar.png}
	\caption{Diffuseness $\psi$ (colormap, solid contour: $90\%$) and level $w$ normalized to origin (dotted \SI{1}{dB} contour) for a 2D, 3:2 ellipse / rectangle ($L_\mathsf{p}$-superellipse with $\mathsf{p}=10$) of 100 equi-angle uncorrelated vertical line sources (salmon): 
    (a,c) with unity gain $\sigma=1$, 
    (b,d) isotropy-enforcing gain $\sigma=\sqrt{R}$. 
	\label{fig:simu3}}
\end{figure*}

Circular arrangements of vertical line sources and spherical arrangements of point sources yield perfect diffuseness $\psi=100\%$ everywhere inside, while sound energy density $w$ varies and slightly increases outwards.

A more disturbing outwards increase of sound energy density is observed either for the circle of point sources as in typical channel-, scene-, or panning-based surround layouts, and even for the circle of horizontal line sources that approximates diffuse-field rendering with 2.5D WFS~\cite{ahrens,sfstoolbox}. Either of these cases seems incapable of rendering of a uniformly high diffuseness, and sound intensity will be dominated by the nearest source.

What seems ideal for circular/spherical source layouts is not ideal for non-circular/non-spherical ellipsoid or superellipse/cuboid layouts. 
\autoref{fig:simu3}~(a) analyzes a 2:3 ellipse of 100 sources arranged evenly in azimuth, whose diffuseness mapping still seems fair, and in (b) the $90\%$-diffuseness contour increases in extent by using a direction-dependent gain that ensures isotropy in the center by inverting the vertical line sources' radial decay by their distance $R$ from there: $\sigma=\sqrt{R}$. 
\autoref{fig:simu3}~(c) analyzes what happens when the equi-angle source arrangement is mapped onto a rectangle, in particular a 2:3 superellipse of the $L_{10}$ norm. There, the result in (c) is worse than with the ellipse (a) with unity gain, and improvement in (d) by  isotropy-enforcing gain $\sigma=\sqrt{R}$ is not as much as for the ellipse. The level (sound energy density) that is plotted in dotted contours remains nearly unaffected by the gain patterns.


When using uncorrelated vertical line sources evenly arranged in a circle as the ideal case, but now discretized, simulation indicates in \autoref{fig:simu2}~(a-d) of \autoref{sec:discretization_study2d} a relative increase of the area within which diffuseness exceeds $\psi\geq90\%$ with the number of loudspeakers.


The observations in the above-mentioned numerical experiments motivate our research questions:
\begin{enumerate}
\item What is the physics behind ideal diffuseness with circular/spherical geometry and matching sources? 
\item Can there be isotropy everywhere inside?
\item Are there optimal sources to produce both, $100\%$ diffuseness and constant sound pressure level?
\item Can results be optimized for non-spherical layouts?
\item What determines the radial limit of discrete layouts?
\item Can 2.5D wave-field synthesis provide diffuseness?
\end{enumerate}

\subsection{Outline}
\autoref{sec:source_distribution} introduces general uncorrelated source distributions in the volume, their statistically expected sound energy density and sound intensity, 
and Gau{\ss}' divergence theorem that would measure the total enclosed variance of all sources, cf.~\cite{Kellogg}. Here, the obvious relation to the solution of Green's functions of potential fields (gravity, electrostatics) is helpful to calculate the particular result based on Green's third identity~\cite{BakerCopson,Green}. 

In particular dealing with \emph{question 1}, Gau{\ss}' divergence can be used to prove that sound intensity must vanish
inside hollow source densities with rotation invariance, or potentially with shift invariance. Such arrangements permit reduction to continuous hulls of sources and can either be a pair of infinite parallel source planes, a thin source shell in the shape of an infinite cylinder, or of a finite sphere, cf.~\autoref{sec:symmetry}. 
\autoref{sec:subspace} clarifies that both correlated and uncorrelated source distributions with infinitely extended excess dimensions exhibit similar radial decays for sound energy density and sound intensity.
Expanding on \emph{question 1} and touching \emph{question 3}, \autoref{sec:newton} shows the formal equivalence to Newton's spherical shell theorem~\cite[prop.70, Sec.12]{Newton}, and explains the underlying geometry to establish a solid understanding of its balance for directionally opposing intensity contributions, for any direction and observer location. 
In $D$ space dimensions, the right balance is accomplished  by a source with the radial decay $\nicefrac{1}{\sqrt{r}^{D-1}}$, whereas the far end would dominate for any smaller exponent, the near end for any higher one.

Concerning \emph{question 2} on isotropy, \autoref{sec:isotropy} uses the integration elements of Newton's spherical shell theorem to observe whether isotropic contributions to sound intensity can be obtained everywhere inside the hull.

To get answers for \emph{question 3}, \autoref{sec:intensitypotential} establishes the scalar potential of sound intensity that becomes constant for vanishing sound intensity. Comparison to the potential of the sound energy density reveals a problematic dimensionality gap for typical physical sources. 

Therefore \autoref{sec:modified_beta} introduces sources of modified decay $\nicefrac{1}{r^\beta}$ more rigorously to study them.
Numerical evaluation will be used to find out which $\beta$ yields ideally large usable listening areas. 
To prove which radial decay exponent $\beta$ is precisely required for vanishing sound intensity or constant sound energy density, \autoref{sec:gegenbauer} expands the modified Green's functions into Gegenbauer polynomials~\cite[15.12.E4]{Gegenbauer77,dlmf} by their generating function. 
Orthogonality will be used to show that constant sound energy density level can be synthesized by an exponent $\beta$ that is either trivial $0$ or smaller by $\nicefrac{1}{2}$ than the exponent delivering a constant sound intensity potential.

A more intuitive proof of the ideal $\beta$ could be successful as well based on Newton's shell theorem alone, or based on inspection of the properties of hypergeometric functions~\cite{Kummer,Goursat}. However, the sophistic and strict proof based on Gegenbauer polynomials also permits to get answers about \emph{question 4} in 
\autoref{sec:suboptimal}:
\autoref{sec:ellipsoid} addresses multi-axial ellipsoids with the direction-dependent radius $R$ seen from their center. In particular, different gain patterns $\sigma$ depending on $R$ will be shown to either provide isotropy for a central observer (same level from all directions) or accomplish perfect synthesis of diffuseness everywhere inside. As more general shapes tunable from ellipsoidal to rectangular cuboid,  \autoref{sec:superellipse} will discuss multi-axis $L_\mathsf{p}$-norm shells. Targeting ideal diffuseness will be shown to work in exemplary layouts by a zeroth-order mode matching technique proposed in \autoref{sec:modematch2} and \autoref{sec:modematch3}. For non-circular/non-spherical arrangements of suitable sources, the corresponding Gegenbauer expansion is re-formulated to an equivalent expansion using circular/spherical harmonics combined with the radial solution of the potential equation.
In both cases, mode matching will be shown to produce suitable amplitude patterns $\sigma$ supporting perfect synthesis of diffuseness inside, however with a negative effect on isotropy. 
As important practical case for 3D audio, \autoref{sec:hemisphere} is dedicated to analyzing upper hemispherical source layouts by numerical simulation, in order to describe how much sound intensity needs to be accepted that vertically traverses the horizontal plane, and how intensity develops when coming closer to the top.

Gegenbauer expansion will also be employed in \autoref{sec:discrete} to address \emph{question 5} and define the limitations of diffusneess synthesis with discrete source layouts in terms of relative radius. To this end, spherical $t$ designs~\cite{Delsarte} are used as sampling schemes for the circle and sphere, because they optimally discretize Gegenbauer expansions. 

\autoref{sec:wfs} deals with \emph{question 6} about whether 2.5D WFS can provide diffuse-field synthesis. To this end, numerical simulations will be used to discuss differently scaled virtual circular layouts of ideally uncorrelated sources, based on their stationary-phase approximation. 

And finally, \autoref{sec:discussion} will discuss the implications and connections of the presented findings to the psychoacoustically relevant interaural metrics used by literature to understand its related listening-experiment studies.

\section{Distributed, uncorrelated sources}\label{sec:source_distribution}
A sound field $p(\bm x)$ that we aim to generate by diffuse sound field synthesis shall be superimposed from a distribution $q(\bm x)$ exciting the Helmholtz equation, cf.~\cite{Skudrzyk71},
\begin{align}
    (\bigtriangleup+k^2)\,p(\bm x)&=-q(\bm x),\label{eq:inhom}
\end{align}
with the Laplacian denoting the second order Cartesian derivatives $\bigtriangleup=\sum_{i=1}^D\frac{\partial^2}{\partial x_i^2}$ for all coordinates $x_i$, and $k=\frac{\omega}{c}$ is the wave number consisting of the frequency argument $\omega=2\pi\,f$  and the speed of sound $c$.

Typically, the inhomogeneous problem is solved via the Green's function of the free field, cf.~\cite{Skudrzyk71},
\begin{align}
    (\bigtriangleup+k^2)\,G(r)&=-\delta(\bm x-\bm x_\mathrm{s}),\label{eq:green}
\end{align}
which is the wave of an excitation at a single source point $\bm x_\mathrm{s}$ to which $r=\|\bm x-\bm x_\mathrm{s}\|$ is the Euclidean distance. This excitation is expressed by the Dirac delta that is zero everywhere else than $\bm x=\bm x_\mathrm{s}$, is normalized so that $\int_V\delta(\bm x-\bm x_\mathrm{s})\,\mathrm{d}V=1$, and fulfills a sifting property $\int_V q(\bm x)\delta(\bm x-\bm x_\mathrm{s})\,\mathrm{d}V=q(\bm x_\mathrm{s})$. For a free sound field, the Green's function must obey Sommerfeld's radiation condition $\lim_{r\rightarrow\infty}\sqrt{r}^\mathrm{D-1}(\mathrm{i}k+\frac{\partial}{\partial r})G=0$~\cite[§28]{Sommerfeld_engl} so that its far-field/high-frequency expression $kr\gg1$ that we are interested in yields, see derivations according to~\cite[Ch.V(A.IV,C),E13]{Sommerfeld_engl} in~\autoref{apdx:hf_green}, eq.~\eqref{eq:hfgreen},
\begin{flalign}
    G&=\sqrt{\frac{2\pi\,r}{\mathrm{i}k}}^{3-D}\frac{e^{-\mathrm{i}kr}}{4\pi\,r}
    =A(k)\,\frac{e^{-\mathrm{i}kr}}{\sqrt{r}^{D-1}}.
    \label{eq:green_hf}
\end{flalign}
The distance-independent term $A(k)$ is typically removed by equalization, and $p(\bm x)$ is obtained by superimposing Green's functions $G$ weighted by $q(\bm x_\mathrm{s})$ for every source point, implying $q(\bm x)=\int_V q(\bm x_\mathrm{s})\,\delta(\bm x-\bm x_\mathrm{s})\,\mathrm{d}V$
as right-hand side of $(\bigtriangleup+k^2)p=-q$, implying the solution for $p$ on the left 
\begin{align}
   p&=\int_V q(\bm x_\mathrm{s})\,G(r)\,\mathrm{d}V.
\end{align}
The sound particle velocity is $\bm v=\frac{\nabla p}{-\mathrm{i}\omega\,\rho}$, and accordingly 
\begin{align}
    \bm v&=\int_V\,a(\bm x_\mathrm{s})\,{\textstyle\frac{\nabla G(r)}{-\mathrm{i}\omega\,\rho}}\,\mathrm{d}V=
\int_V\,a(\bm x_\mathrm{s})\,{\textstyle\frac{\bm u\,G(r)}{-\mathrm{i}\omega\,\rho}}\,\mathrm{d}V,
\end{align}
where the Green's function $G\propto b\,e^{-\mathrm{i}k\,r}$ was approximated for the far field/high frequencies $kr\gg1$ by
\begin{align}
    \nabla G(r)&=(\nabla r)(G'(r))=
    \bm u\,[{\textstyle\frac{1}{b}\frac{\partial b}{\partial r}-\mathrm{i}k}]\,G\nonumber\\
 &\approx -\mathrm{i}k\,\bm u\,G,
\end{align}
and $\bm u=\frac{\bm x-\bm x_\mathrm{s}}{\|\bm x-\bm x_\mathrm{s}\|}$ is the direction vector to the receiver.

\begin{figure*}[t]
    \centering    \includegraphics[width=1.0\textwidth]{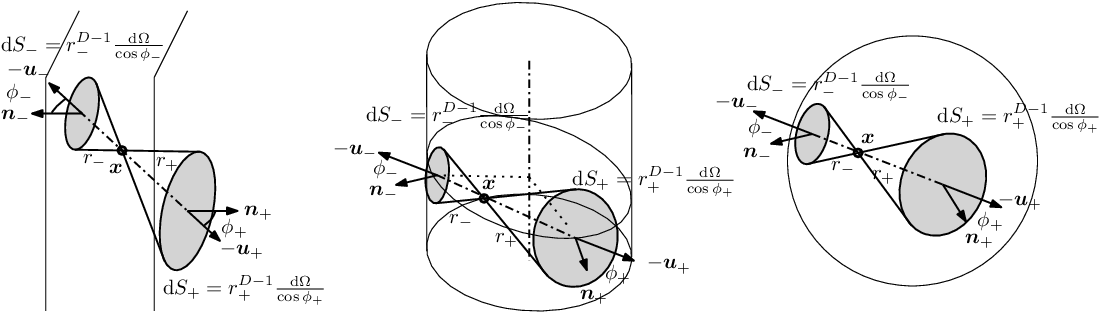}
    \caption{Differential surface elements in $\mathbb{R}^3$ for symmetric arrangements of (left) infinite parallel planes with shift invariance in $yz$, (middle) an infinite cylinder with and rotational symmetry in $xy$ and shift invariance in $z$, and (right) a sphere with rotational symmetry in $xyz$, according to Newton's spherical shell theorem.}
    \label{fig:newton123}
\end{figure*}
We assume an amplitude distribution $q(\bm x_\mathrm{s})$ of uncorrelated/incoherent sources of the variance $\sigma^2$ in the integration volume
\begin{align}
    E\{q(\bm x_1)^* q(\bm x_2)\}=\sigma^2\,\delta(\bm x_1-\bm x_2)
\end{align} 
with the statistical expectation operator denoted as $E\{\}$.
The Dirac delta is zero unless $\bm x_1=\bm x_2$ and normalized.

\subsection{Expected sound energy density, sound intensity, sound power}
\label{sec:expectation}
The sound energy density $\rho\,c^2\,w=E\{|p|^2\}$ scaled by $\rho\,c^2$ equals the expected mangitude-square sound pressure, cf.~\cite{Kuttruff},
\begin{align}
    \rho\,c^2\,w&=\iint_V\,E\{q(\bm x_1)^*q(\bm x_2)\}\,G_1^*G_2\,\mathrm{d}V_1\mathrm{d}V_2\nonumber\\
&=
\int_V\,\sigma^2\,|G(r)|^2\,\mathrm{d}V,\label{eq:w_vol}
\end{align}
and sound intensity $\rho\,c\,\bm I=E\{p^*\bm v\}$ times  $\rho\,c$ is, cf.~\cite{Kuttruff}, 
\begin{align}
    \rho\,c\,\bm I&=\iint_V\,E\{q(\bm x_1)^*q(\bm x_2)\}\,\bm u\,G_1^*G_2\,\mathrm{d}V_1\mathrm{d}V_2\nonumber\\
&=
\int_V\,\sigma^2\,\bm u\,|G(r)|^2\,\mathrm{d}V.\label{eq:I_vol}
\end{align}
\paragraph{Gau{\ss}' law} permits to estimate the sound power $\dot W$ of sources enclosed in a hull $S=\partial V$ via net intensity flux out of the hull, or the divergence $\nabla^\intercal \bm I=:\dot w$, the sound power density, integrated over the enclosed volume $V$
\begin{align}
    \dot W=\rho\,c\,\Re\Big\{\int_S\bm I^\intercal\,\bm n\,\mathrm{d}S\Big\}&=
    \rho\,c\,\Re\Big\{\int_V\nabla^\intercal \bm I\,\mathrm{d}V\Big\}.\label{eq:gausslaw}
\end{align}
It is helpful to notice that squared free-field Green's function of the Helmholtz equation times direction 
\begin{align}
  \bm u\,|G(r)|^2&=\frac{\bm u}{\,r^{D-1}} {|A|^2}=-\mathcal{S}_{D-1}\,|A|^2\,\nabla\mathcal{G}(r)\\
  &=\mathcal{S}_{D-1}\,|A|^2\,\bm u\,\mathcal{G}'(r)\nonumber
  ,\label{eq:grad_green_laplace}
\end{align}
relates to the Green's function $\mathcal{G}$ of the Laplace or potential equation (and $\mathcal{S}_{D-1}\,\mathcal{G}'=-\frac{1}{r^{D-1}}$)
\begin{align}
    \bigtriangleup\,\mathcal{G}(r)&=-\delta(\bm x-\bm x_\mathrm{s})\nonumber\\
	\mathcal{G}(r)&=\begin{cases}
		-\frac{1}{\mathcal{S}_{D-1}}\ln r & D=2,\\
		\frac{1}{(D-2)\,\mathcal{S}_{D-1}}\frac{1}{r^{D-2}} & D\neq2,
	\end{cases}
\end{align}
and $\mathcal{S}_{D-1}=\frac{2\,\sqrt{\pi}^{\mathrm{D}}}{\Gamma(\frac{\mathrm{D}}{2})}$ is the area of the unit sphere in $D$ dimensions \cite[vol.2, p.387]{HilbertCourant}.
Divergence of gradient $\nabla^\intercal(\nabla\mathcal{G})=\bigtriangleup\mathcal{G}=-\delta$ yields an integrand of eq.~\eqref{eq:gausslaw} 
\begin{align}
  \rho\,c\,\nabla^\intercal\bm I&=\int_V\sigma^2(\bm x_\mathrm{s})\,\delta(\bm x-\bm x_\mathrm{s})\,\mathcal{S}_{D-1}\,|A|^2\,\mathrm{d}V\nonumber\\
  &=\mathcal{S}_{D-1}\,|A|^2\,\sigma^2(\bm x),
\end{align}
and upon insertion, the total sound power of all sources enclosed in $V$ using Gau{\ss}' law eq.~\eqref{eq:gausslaw} becomes
\begin{align}
    \int_S\bm I^\intercal\bm n\,\mathrm{d}S=\mathcal{S}_{D-1}\,|A|^2\,\int_V\sigma^2\,\mathrm{d}V=\dot W.
\end{align}
In this article, we will work with a volume $V$ that is free of sources inside, so that sound power $\dot W$ vanishes inside
\begin{align}
   \dot W&=\int_S\bm I^\intercal\bm n\,\mathrm{d}S=0, &\text{as }\sigma(\bm x)&=0\;\forall\bm x\in V.\label{eq:sourcefree}
\end{align}
In particular, we continue after re-defining the volume source distribution to a thin shell $S$ of sources distributed tightly around $V$, just not contained within $V$ anymore, so that the interior is source-free. Adapting eqs.~\eqref{eq:w_vol} \eqref{eq:I_vol}, these shell integrals become
\begin{align}
   \rho\,c^2\,w&=\frac{1}{\hat N}\int_S \sigma^2\,|G(r)|^2\,\mathrm{d}S,\label{eq:w}
   \\
   \rho\,c\,\bm I&=\frac{1}{\hat N}\int_S \sigma^2\,\bm u\,|G(r)|^2\,\mathrm{d}S,\label{eq:I}
\end{align}
where $\hat N$ is used to normalize 
$\rho\,c^2\,w(\bm 0)=1$.
Around the line between any surface point $\bm{x}_\mathrm{s}$ and the observer $\bm x$, defining the direction $\bm u=\frac{\bm x-\bm x_\mathrm{s}}{\|\bm x-\bm x_\mathrm{s}\|}$, the surface element $\mathrm{d}S$ is enclosed in an infinitesimally small angular cone $\mathrm{d}\Omega$ intersecting with the surface normal $\bm n$ at an angle $\phi$, and it scales by projection $\cos\phi=\bm n^\intercal\bm u$ and a distance-related area increase $r^{D-1}$ so that  $\mathrm{d}S=r^{D-1}\,\frac{\mathrm{d}\Omega}{\cos\phi}$.

Note that while $\mathrm{d}S$ is most suitable to cover the fundamentals, we use instead of $\mathrm{d}S(\bm u)$ a modified integral over $\mathrm{d}\Omega_0(\bm u_0)$ from \autoref{sec:intensitypotential} on and in the introductory examples, referring to an angular segment seen in the direction $\bm u_0$ from the origin, yielding a wave towards $\bm u(\bm u_0)$ at the observer.

\section{Optimal hulls of sources}\label{sec:symmetry}
A source-free interior of the volume $V$ as in eq.~\eqref{eq:sourcefree} can be combined with a symmetry assumption that determines the intensity in eq.~\eqref{eq:sourcefree} and forces it to zero. In particular, if the intensity: 
\begin{itemize}
    \item has a constant magnitude at every surface position $\|\bm I\|\neq \|\bm I(\bm x_\mathrm{s})\|$ and
    \item is non-tangential, so strictly aligned with the surface normal for every surface position
    $\bm I=I_n\,\bm n$
\end{itemize}
then intensity vanishes due to the source-free interior ($\dot W=0$) and the non-zero hull area ($\oint_S\mathrm{d}S=S\neq0$) 
\begin{align}
   \dot W=\oint_S\bm I^\intercal\bm n\,\mathrm{d}S&=
   I_n\oint_S\bm n^\intercal\bm n\,\mathrm{d}S=I_n\,S=0,\nonumber\\
    \Rightarrow I_\mathrm{n}\equiv 0\Rightarrow \bm I&\equiv\bm 0\;\forall\bm x\in S.\label{eq:Isymm}
\end{align}
All-zero intensity components everywhere on $S$ entails zero intensity $\bm I\equiv\bm 0\;\forall \bm x\in V$ everywhere in the source-free field enclosed.
Similar considerations are employed in standard literature on electrostatic fields, cf.~\cite{Ling16}.

The two constraints are fulfilled in particular if the variance of the sources is invariant $\sigma^2(\bm x)=\sigma^2(\bm R\,\bm x)$ under rotation by any matrix $\bm R\in\mathbb{R}^{D'}$: $\bm R^\intercal\bm R=\mathbf{I}$, $\mathrm{det}\{\bm R\}=1$. Moreover rotation invariance can be limited to $D'\leq D$ subspace dimensions $D'>0$,
for which one could, e.g., define a spherical manifold of the radius $\sum_{i=1}^{D'}x_{\mathrm{s},i}^2=R^2$. 
Rotation invariance will ensure there being no tangential components $\bm I= I_n\bm n$, as long as there is shift invariance for any shift $\Delta x\in\mathbb{R}^{D-D'}$ along the excess dimensions $i>D'$, 
$\sigma^2(\bm x)=\sigma^2(\bm x+\Delta\bm x)$. 
Considering a single spherical manifold radius $R$ as a shell of sources in $\mathbb{R}^3$, requirements include, cf.~\autoref{fig:newton123}: 
\begin{enumerate}
    \item an opposing pair of infinite planes, e.g., $x_{\mathrm{s},1}^2=R^2$
    \item an infinite cylinder, e.g., $x_{\mathrm{s},1}^2+x_{\mathrm{s},2}^2=R^2$
    \item a finite sphere, $x_{\mathrm{s},1}^2+x_{\mathrm{s},2}^2+x_{\mathrm{s},3}^2=R^2$.
\end{enumerate}

\subsection{Reduced radial decay for infinite extent}\label{sec:subspace}
When there are no infinitely extended excess dimensions $E=D-D'=0$ of the source distribution, the function $\mathcal{G}'(r)\propto\frac{1}{r^{D-1}}$ is responsible for the ideally vanishing intensity in eq.~\eqref{eq:Isymm}. When there are $E>0$ excess dimensions with uncorrelated sources being integrated over, we may use the relations shown in \autoref{apdx:excess}
to find
\begin{align}
\lim_{M\rightarrow\infty}\underbrace{\int_{-M}^{M}\dots\int_{-M}^{M}}_E \frac{\prod_{i=D'+1}^D\mathrm{d}x_{\mathrm{s},i}}{(2L)^E\,r^{D-1}}\propto\frac{1}{r^{D'-1}}.\label{eq:reduced_decay_excess}
\end{align}
This determines the function  $\mathcal{G}'_{D'}\propto\frac{1}{r^{D'-1}}$ for the remaining subspace. Constant multipliers and scaling by the size of the excess surface in the limit towards infinity can be usefully suppressed by re-defining $\sigma^2$ accordingly. 

It is noteworthy that an equivalent  reduction of the exponent occurs when driving all sources  in phase. Here, the infinite extent is  obtained by removing the independent dimension from the Helmholtz equation, yielding a free-field Green's function $G_{D'}\propto\frac{e^{-\mathrm{i}kr}}{\sqrt{r}^{D'-1}}$ for $kr\gg1$, cf.~eq.~\eqref{eq:green_hf}, of which the magnitude square in eq.~\eqref{eq:I} is equivalent $|G_{D'}|^2\propto\mathcal{G}'_{D'}\propto\frac{1}{r^{D'-1}}$.

Only the uncorrelated sources can create incoherence along the excess dimension, but the metrics $w$, $\bm I$, $\psi$ are blind to this aspect.
It is formally equivalent when either the infinite dimensions stem from a cylinder-wave or plane-wave radiator, or they stem from a linear or planar distribution of uncorrelated point sources, respectively. 
Below, we simplify $D'$ to $D$ and skip explicit assumptions about correlation along excess dimensions.

\subsection{Newton's spherical shell theorem}
\label{sec:newton}
Newton's spherical shell theorem~\cite[prop.70, Sec.12]{Newton}, concludes that the gravitational force due to a thin hollow spherical shell vanishes inside. 
The contribution $\mathrm{d}F$ of a surface element $\mathrm{d}S$ centered at $\bm x_\mathrm{s}$ to the force $\bm F=\int\bm u\,\mathrm{d}F$ at $\bm x$ is proportional to the factor $\frac{1}{r^{D-1}}$ (inverse square law for $D=3$ of $\mathbb{R}^3$) of its distance $r=\|\bm x-\bm x_\mathrm{s}\|$ times $\mathrm{d}S=r^{D-1}\,\frac{\mathrm{d}\Omega}{\cos\phi}$, related to the angular element $\mathrm{d}\Omega$ as  described above.
The force contributions $\mathrm{d}F_-$ and $\mathrm{d}F_+$ of opposing intersections along  $\pm\bm u$ annihilate 
\begin{align}
   \frac{r_+^{D-1}}{r_+^{D-1}}\,\frac{\mathrm{d}\Omega}{\cos\phi_+}=
   \frac{r_-^{D-1}}{r_-^{D-1}}\,\frac{\mathrm{d}\Omega}{\cos\phi_-},\label{eq:newton}
\end{align}
i.e.\ whenever both intersection angles match $\phi_+=\phi_-$. This is obviously the case for parallel dimensions. Also, any opposing pair of intersection points $\bm x_\mathrm{s}^+$ and $\bm x_\mathrm{s}^-$ with a spherical manifold ensures $\phi_+=\phi_-$ by the isosceles triangle enclosed with the manifold's origin $x_{i}=0\,\forall i\leq D'$, cf.~\autoref{fig:newton123}~(middle). With this match for any $\pm\bm u$, integration over all directions $\bm u$ yields a vanishing net force $\bm F=\bm 0$.
Despite a global sign difference, $\mathrm{d}\bm I=-\mathrm{d}\bm F$, the formulation for sound intensity is equivalent.

Assuming $r_+>r_-$ and symmetry $\phi_+=\phi_-$, Newton's theorem shows that imbalance can arise from a modified exponent $\nicefrac{1}{r^\alpha}$, cf.~\autoref{fig:wavefronts_cc},  between near-end $\mathrm{d}I_-=\frac{r_-^{D-1}}{r_-^{\alpha}}\,\frac{\mathrm{d}\Omega}{\cos\phi_-}$ and far-end $\mathrm{d}I_+=\frac{r_+^{D-1}}{r_+^{\alpha}}\,\frac{\mathrm{d}\Omega}{\cos\phi_+}$ whenever $\alpha\neq D-1$: 
An exponent that is too high causes the near end to dominate intensity, one that is too small causes dominance of the far end, cf.~\autoref{sec:modified_beta}.

\subsection{Isotropic directional intensity}
\label{sec:isotropy}
Nolan et al.~\cite{nolan} expressed isotropy as the uniformity of the wave-number magnitude spectrum, i.e.\ the uniform magnitude for all plane-wave directions received. Equivalently corresponding with the definitions used for the shell theorem, we may define isotropy via the intensity contribution from an infinitesimal angular cone $\mathrm{d}\Omega$ aligned with a variable direction of arrival $-\bm u$,
\begin{align}
   \mathrm{d}I_\mathrm{u}&=
   \frac{\sigma^2}{r^{D-1}}\,\frac{r^{D-1}\mathrm{d}\Omega}{\cos\phi}=\sigma^2\,\frac{\mathrm{d}\Omega}{\cos\phi},
\end{align}
so that isotropy can be measured by how constant
\begin{align}
   \frac{\mathrm{d}I_\mathrm{u}}{\mathrm{d}\Omega}=\frac{\sigma^2}{\cos\phi}=\text{const}
\end{align}
evolves across directions $\bm u$. This is a more strict constraint than Newton's spherical shell theorem that just requires for every direction $\bm u$ the annihilation with the opposing direction $-\bm u$ by $\frac{\mathrm{d}I_\mathrm{u}}{\mathrm{d}\Omega}=\frac{\mathrm{d}I_\mathrm{-u}}{\mathrm{d}\Omega}$. The strictness of isotropy lies in direction-independence, which is only accomplished (i) at the center of a perfectly spherical layout, $D'=D$, $\sigma=1$, and  $r=R=\text{const}$, where $\cos\phi\equiv 1$, or (ii) at a particular point of observation within an arbitrary convex layout for which the source variance is adjusted to
\begin{align}
   \sigma^2\propto\cos\phi,
\end{align}
for every direction $\bm u$.
Assuming to accomplish perfect isotropy by this fixed $\sigma^2(\bm u_\mathrm{s})$ that matches the respective direction cosine $\cos\phi_0$  at every source to a central observer at $\bm x=\bm 0$, an observer shifted off center will yield a different direction cosine $\cos\phi$ at each source, in general. This contradicts a fixed choice of $\sigma^2$, which can only enforce isotropy for a specific point inside $V$. We will therefore restrict our subsequent discussion of isotropy to a central observer in \autoref{sec:ellipsoid} and \autoref{sec:superellipse}.

\subsection{Intensity potential, energy density}\label{sec:intensitypotential}
For a thin source shell in the subspace $\mathbb{R}^{D}$, we fix the variance to a constant value $\sigma^2$ that is non-zero only at a fixed radius, e.g.\ $R=1$ of the unit sphere, denoted as $\mathbb{S}$. According to eq.~\eqref{eq:Isymm}, symmetry ensures vanishing sound intensity $\bm I=0$ inside $\bm x\in V$
\begin{align}
   \rho\,c\,\bm I&= \frac{\sigma^2}{\tilde N}\,\int_{\mathbb{S}}\nabla^\intercal\mathcal{G}_{D}(r)\,\mathrm{d}\Omega_0=\bm 0.\label{eq:I_calG}
\end{align}
Here, we used a surface element 
$\mathrm{d}S=
\mathcal{S}_{D-2}\,R^{D-1}\,\mathrm{d}\Omega_0$ of an isotropic angular segment $\mathrm{d}\Omega_0$ seen from the origin, the area of the unit sphere 
$\mathcal{S}_{D-2}$ reduced by a
dimension (the area of the unit sphere is $\mathcal{S}_{D-1}=\frac{2\,\sqrt{\pi}^{\mathrm{D}}}{\Gamma(\frac{\mathrm{D}}{2})}$, cf.~\cite[vol.2, p.387]{HilbertCourant}),  a constant source variance 
$\sigma^2=\text{const}$, and a new normalizer $\tilde N$ to hide some of the constant multipliers. To relate Laplacian and Helmholtz Green's functions, eq.~\eqref{eq:grad_green_laplace}, we used
\begin{align}
    \mathcal{S}_{D-1}|A|^2\;\mathcal{G}'_{D}(r)=|G_{D}(r)|^2.
\end{align}

\begin{figure*}[t]
    \centering    \includegraphics[width=17cm,trim=8mm 1.2cm 4mm 9.9cm,clip]{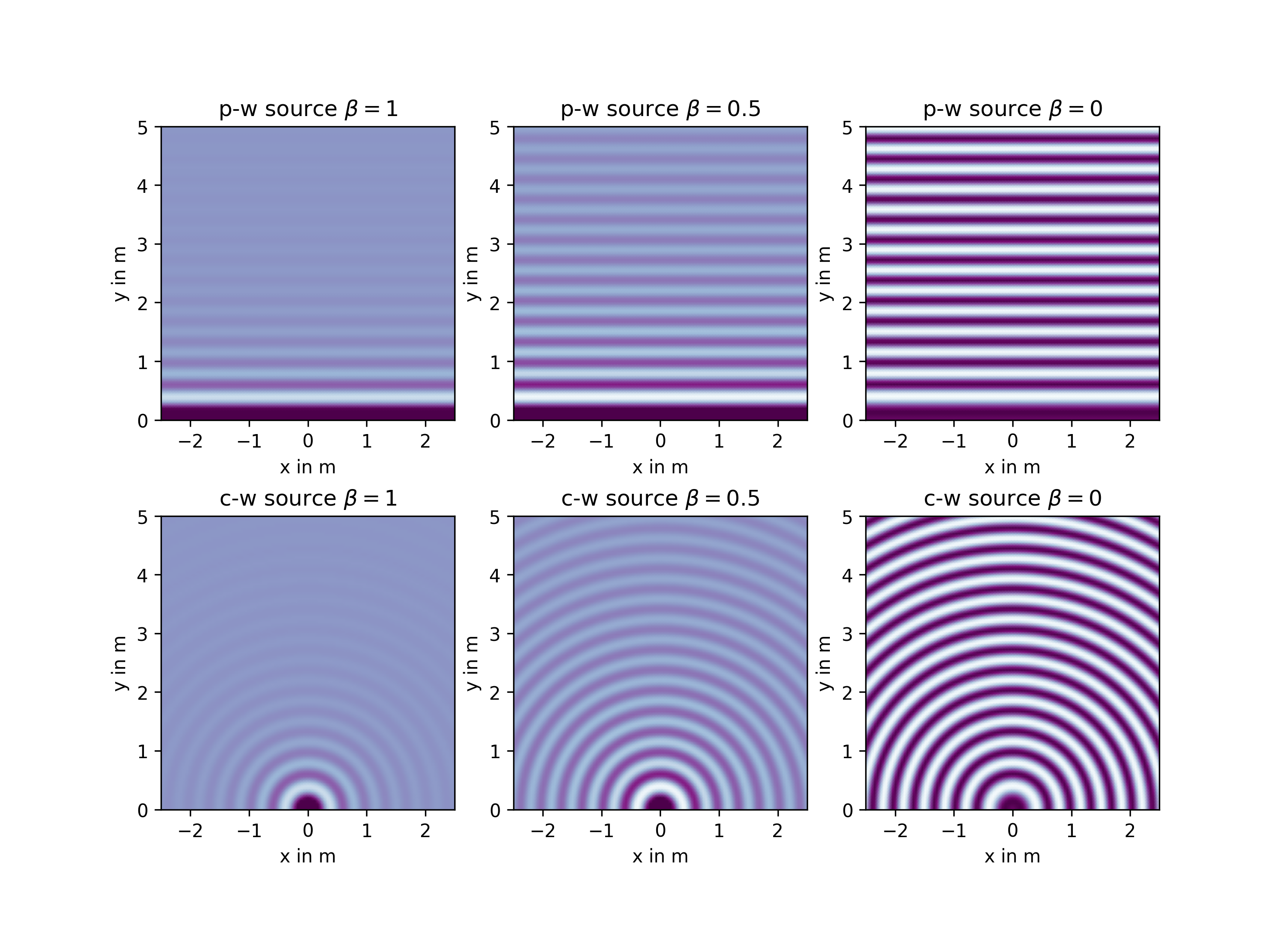}
    \caption{Point source at $\bm x_\mathrm{s}=\bm 0$ with radial decay exponent modified from left to right $\beta=\{1,0.5,0\}$, at $f=\SI{1}{kHz}$. }
    \label{fig:wavefronts_cc}
\end{figure*}

We can conceive a potential field $-U_\mathrm{I}$ whose gradient delivers the intensity $-\nabla U_\mathrm{I}=\rho\,c\,\bm I$. Because $\bm I=\bm 0$, the related potential field is spatially constant
\begin{align}
-U_\mathrm{I}&=\frac{\sigma^2}{\tilde N}\int_{\mathbb{S}}\mathcal{G}_{D}(r)\,\mathrm{d}\Omega_0=\text{const}.
\end{align}
While for the synthesis of vanishing sound intensity, eq.~\eqref{eq:Isymm} and Newton's spherical shell theorem eq.~\eqref{eq:newton} are proof that spherical shells describe the ideal geometry,  
these relations may indicate that sound-energy density, based on  $\mathcal{G}'$ instead of $\mathcal{G}$, is not constant,
\begin{align}
   \rho\,c^2\,w&=
   \frac{\sigma^2}{\tilde N}\int_{\mathbb{S}}\mathcal{G}'_{D}(r)\,\mathrm{d}\Omega_0\neq\text{const?},\label{eq:w_calG}
\end{align}
unless $D=1$, i.e., a shell of two infinite planes, which is a trivial case; $\tilde{N}$ normalizes $\rho\,c^2\,w(\bm0)=1$ for $\mathcal{G}_D'$.

\section{Source of modified decay $\beta$}\label{sec:modified_beta}
\begin{figure*}[t]
	\centering
	\subfigure[sound energy density for circular (left) / spherical (right) layout\label{fig:betas_w}]{\includegraphics[height=3.0cm,trim=0mm 0 0 9mm,clip]{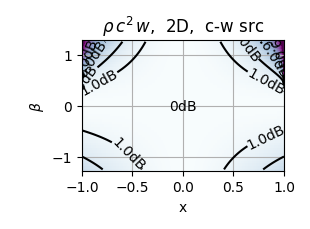}
    \includegraphics[height=3.0cm,trim=11mm 0 0 9mm,clip]{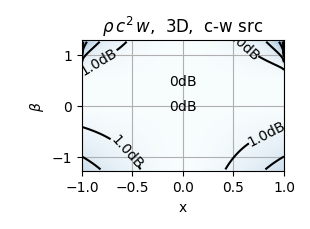}}
    \subfigure[diffuseness for circular (left) / spherical (right) layout\label{fig:betas_I}]{
    \includegraphics[height=3.0cm,trim=13mm 0 0 9mm,clip]{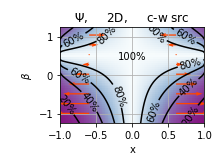}
    \includegraphics[height=3.0cm,trim=13mm 0 0 9mm,clip]{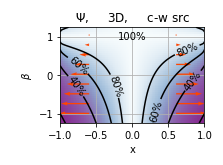}}
	\caption{Evaluation of sound energy density (a) and diffuseness (b) for $-1.25\leq\beta\leq1.25$ (vertical axis) within $-1\leq x\leq 1$ (horizontal axis; orange-red arrows display $\nicefrac{I_\mathrm{x}}{c\,w}$ at $x=\pm 0.6$).\label{fig:betas}}
 \vspace{-2mm}
\end{figure*}

To find conditions for constant sound energy density, we test a free-field Green's function with manipulated radial decay, cf.~\autoref{fig:wavefronts_cc}, for which practical implementation in 2D layouts was suggested
using curved/phased vertical line-source arrays~\cite{straube2018an,Goelles},
\begin{align}
    G_\beta(r)&=\frac{e^{-\mathrm{i}kr}}{r^\beta},
    &
    |G_\beta(r)|^2&=\frac{1}{r^{2\beta}}=:\mathcal{G}_{2\beta}'(r),
\end{align}
and it modifies the related Laplacian Green's function to
\begin{align}
    \mathcal{G}_{2\beta}(r)&=\begin{cases}
        \frac{1}{-(2\beta-1)r^{2\beta-1}}, &2\beta\neq1,\\
        \ln r, &2\beta=1.
    \end{cases}
\end{align}
We test a normalized displacement $-1\leq x\leq 1$ for the coordinate $x_1=x$ and $x_i=0$ else for the shell radius $R=1$ of a unit sphere.
Integrals for $\rho\,c^2\,w$, $\rho\,c\,\bm I$, $\psi$ use the angular fraction $\frac{\mathrm{d}\Omega_0}{\mathcal{S}_{D-1}}=\frac{\mathcal{S}_{D-2}}{\mathcal{S}_{D-1}}\sqrt{1-z^2}^{D-3}\mathrm{d}z$ related to the direction cosine $z=\cos\phi$ between source and observer directions from the origin, the unit-sphere areas $\mathcal{S}_{D-1}$, $\mathcal{S}_{D-2}$ \cite[vol.2, p.387]{HilbertCourant}, the weight
$\sqrt{1-z^2}^{D-3}$, for generalization\footnote{Altogether, this ensures $\frac{\mathrm{d}\Omega_0}{\mathcal{S}_{D-1}}=\frac{2\pi\,\mathrm{d}\cos\phi}{4\pi}$ for $D=3$ and $\frac{\mathcal{S}_{D-2}}{\mathcal{S}_{D-1}}\sqrt{1-z^2}^{D-3}\mathrm{d}z=\frac{\mathrm{d}\phi}{2\pi}$ for $D=2$, cf.~\cite{HilbertCourant}.}, and the corresponding distance $r=\sqrt{1-2zx+x^2}$ between source and observer.
\autoref{apdx:hypergeometric} describes the integrals in greater detail and substitutes $z=2t-1$ within $-1\leq z\leq 1$ with the variable $t$ within $0\leq t\leq 1$ to make them fit the integral representation \cite[15.6.E1]{dlmf},
\begin{align}
    F(a,b;c;z)&=\frac{\Gamma(c)}{\Gamma(b)\Gamma(c-b)}\int_{0}^1\frac{t^{b-1}\,(1-t)^{c-b-1}}{(1-zt)^a}\;\mathrm{d}t,\nonumber
\end{align}
of Gau{\ss}' hypergeometric function $F(a,b;c;z)$ \cite[15.2.E1]{dlmf}.
Sound energy density $\rho\,c^2\,w$ and the only non-zero sound intensity component $\rho\,c\,I_\mathrm{x}$ along $x_1$ become
\begin{align}
\rho\,c^2\,w=&\frac{F\textstyle(\beta,\frac{D-1}{2};D-1;\frac{4x}{(1+x)^2})}{(1+x)^{2\beta}},\label{eq:w_beta}\\
\rho\,c\,I_\mathrm{x}=&
\frac{F\textstyle(\beta+\frac{1}{2},\frac{D-1}{2};D-1;\frac{4x}{(1+x)^2})}{(1+x)^{2\beta}}-
 \nonumber\\
&
\qquad\frac{F\textstyle(\beta+\frac{1}{2},\frac{D+1}{2};D;\frac{4x}{(1+x)^2})}{(1+x)^{2\beta+1}},
\label{eq:I_beta}
\end{align}
and correspondingly diffuseness becomes
\begin{align}
\psi=&1-\bigg|\frac{F\textstyle(\beta+\frac{1}{2},\frac{D-1}{2};D-1;\frac{4x}{(1+x)^2})}{F\textstyle(\beta,\frac{D-1}{2};D-1;\frac{4x}{(1+x)^2})}\nonumber\\&\qquad-\frac{\frac{1}{1+x}F\textstyle(\beta+\frac{1}{2},\frac{D+1}{2};D;\frac{4x}{(1+x)^2})}{F\textstyle(\beta,\frac{D-1}{2};D-1;\frac{4x}{(1+x)^2})}\bigg|.
\label{eq:psi_beta}
\end{align}
Python implements the hypergeometric function as \verb|scipy.special.hyp2f1|, and results  were verified by numerical integrals.
\autoref{fig:betas_w} displays the resulting sound energy density $\rho\,c^2\,w$ eq.~\eqref{eq:w_beta} and \autoref{fig:betas_I} the diffuseness $\psi$ eq.~\eqref{eq:psi_beta} contours for the displacements $x$ normalized to lie between $-1\leq x\leq 1$ and $\beta$ varied between $-1.25\leq\beta\leq1.25$, either for a 2D circle arrangement of sources or a 3D sphere arrangement of sources. At the center $x=0$, ideal values are always reached. In the circular arrangement $D=2$, an extended spatial range requires $\beta=0$ for constant sound energy density, so unphysical point sources without decay, while constant $100\%$ diffuseness requires $\beta=\frac{1}{2}$, so physical vertical line sources. For the spherical arrangement $D=3$, constant sound energy density requires $\beta=\{0,\frac{1}{2}\}$, so either unphysical point sources without decay or unphysical ones with the decay of a line source, while $100\%$ diffuseness requires physical point sources with $\beta=1$. 

The sound energy density seems to vary only moderately with non-ideal, but physical values for $\beta$ of the given $D$. It appears reasonable to focus on accomplishing $100\%$ diffuseness with physical sources while accepting a  sound energy density that is not perfectly constant.

\subsection{Optimal $\beta$ via Gegenbauer polynomials}\label{sec:gegenbauer}
As a strict proof of the optimal $\beta$, we regard the generating function of the orthogonal Gegenbauer polynomials~\cite[18.12.E4]{dlmf} that permits to expand our modified Green's functions in $r=\sqrt{1-2zx+x^2}$ as
\begin{align}
    \mathcal{G}'_{2\beta}(r)\propto\frac{1}{r^{2\beta}}=&\sum_{n=0}^\infty x^n\,C_n^{(\beta)}(z),\\
    \mathcal{G}_{2\beta}(r)\propto\frac{1}{r^{2\beta-1}}=&\sum_{n=0}^\infty x^n\,C_n^{(\beta-\frac{1}{2})}(z),
\end{align} 
where $\mathcal{G}'_{2\beta}$ is integrated for $\rho\,c^2\,w$, cf.~eq.~\eqref{eq:w_calG} and
$\mathcal{G}_{2\beta}$ for $U_\mathrm{I}$, cf.~eq.~\eqref{eq:I_calG}.
As above, integration of these modified Green's functions over the shell uses the axisymmetric surface element $\mathcal{S}_{D-2}\,\varpi_{\alpha}\,\mathrm{d}z$ for integration between $-1\leq z\leq 1$ along the direction of the shift $x$, and the surface weight $\varpi_{\alpha}(z)=\sqrt{1-z^2}^{2\alpha-1}$, cf.~\cite{HilbertCourant}. It matches the manifold dimensions when $2\alpha=D-2$. We choose a constant variance $\sigma^2$ to eliminate independent scalars in the sound energy density and sound intensity potential
\begin{align}
   \rho\,c^2\,w&=\sum_{n=0}^\infty x^n\,\int_{-1}^{1}
   C_n^{(\beta)}(z)\,\varpi_{\alpha}(z)\,\mathrm{d}z,\label{eq:wpot}\\
   -U_\mathrm{I}&\propto\sum_{n=0}^\infty x^n\,\int_{-1}^{1}
   C_n^{(\beta-\frac{1}{2})}(z)\,\varpi_{\alpha}(z)\,\mathrm{d}z\label{eq:ipot}
   .
\end{align}
As the zeroth degree Gegenbauer polynomial is both constant $C_0^{(\nu)}(z)=1$ and orthogonal to any higher-degree Gegenbauer polynomial of the specific family $\nu$, integrals for $n>0$ must vanish $\int_{-1}^1C_n^{(\nu)}(z)\,\varpi_{\nu}\,\mathrm{d}z=0$, as long as the surface weight $\varpi_\nu$ matches the family $\nu$.

\begin{figure*}[t]
	\centering
    	\subfigure[ellipse $\sigma=1$]{
 \includegraphics[height=37mm,trim=4mm 4mm 3mm 3mm,clip]{ellipse_L100_vert_line_srcs_nocomp.png}}	
 \subfigure[ellipse $\sigma=\sqrt{R}$]{
 \includegraphics[height=37mm,trim=4mm 4mm 3mm 3mm,clip]{ellipse_L100_vert_line_srcs_comp.png}}
  \subfigure[ellipse $\sigma=R$]{
 \includegraphics[height=37mm,trim=4mm 4mm 3mm 3mm,clip]{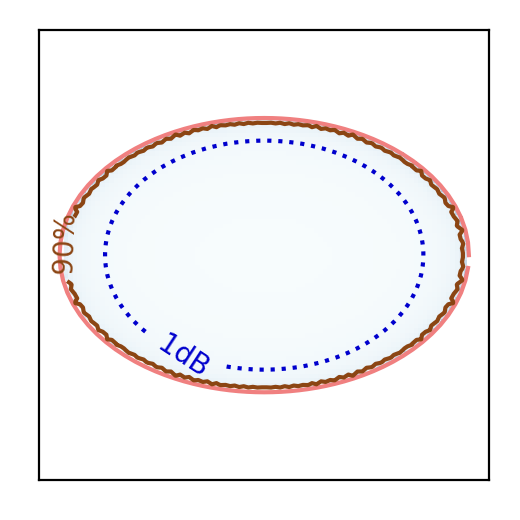}}
   \subfigure[CH mode-match $\sigma$]{
 \includegraphics[height=37mm,trim=4mm 4mm 3mm 3mm,clip]{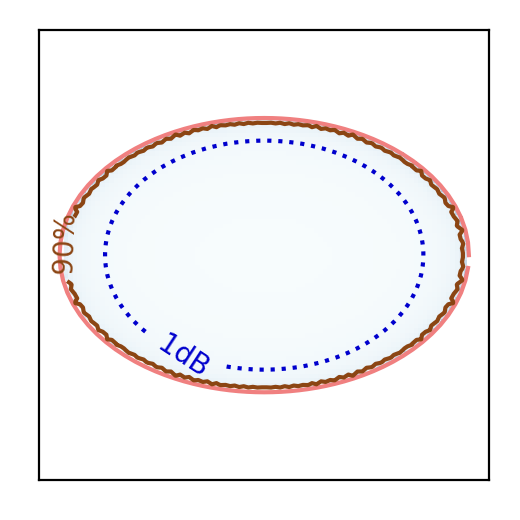}}
 \quad\includegraphics[height=38mm,trim=0mm 5mm 0mm 5mm,clip]{colorbar.png}

     	\subfigure[ellipsoid $\sigma=1$]{
 \includegraphics[height=37mm,trim=4mm 4mm 3mm 3mm,clip]{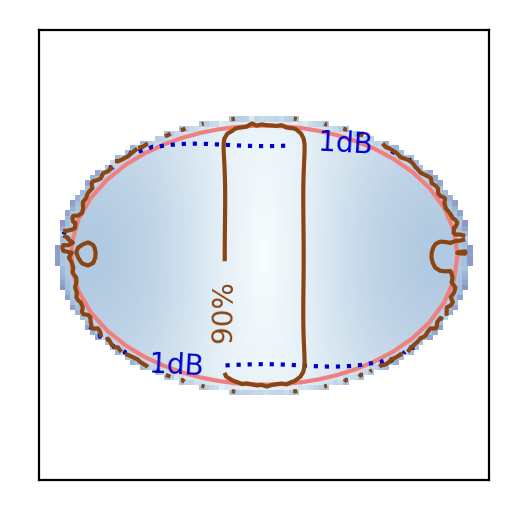}}	
 \subfigure[ellipsoid $\sigma={R}$]{
 \includegraphics[height=37mm,trim=4mm 4mm 3mm 3mm,clip]{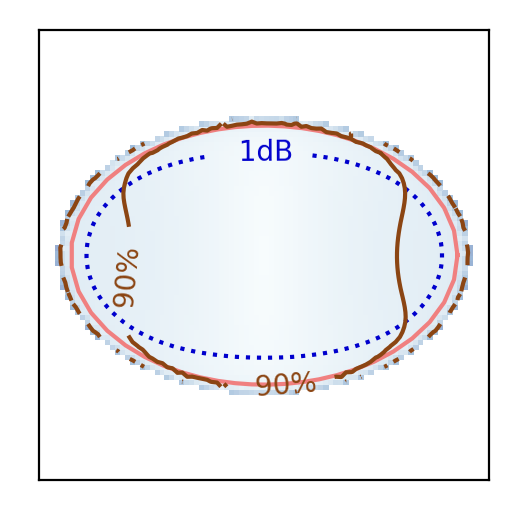}}
  \subfigure[ellipsoid $\sigma=\sqrt{R}^3$]{
 \includegraphics[height=37mm,trim=4mm 4mm 3mm 3mm,clip]{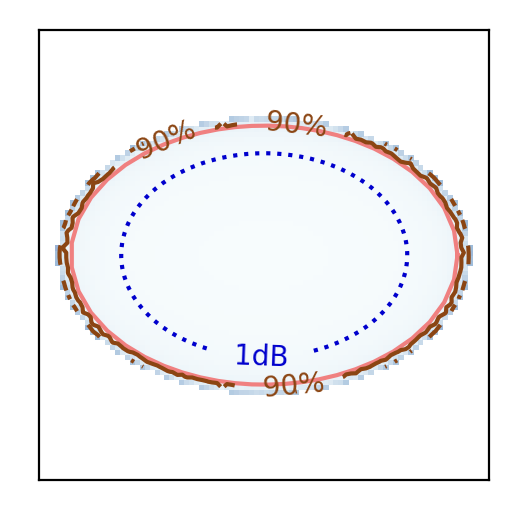}}
   \subfigure[SH mode-match $\sigma$]{
 \includegraphics[height=37mm,trim=4mm 4mm 3mm 3mm,clip]{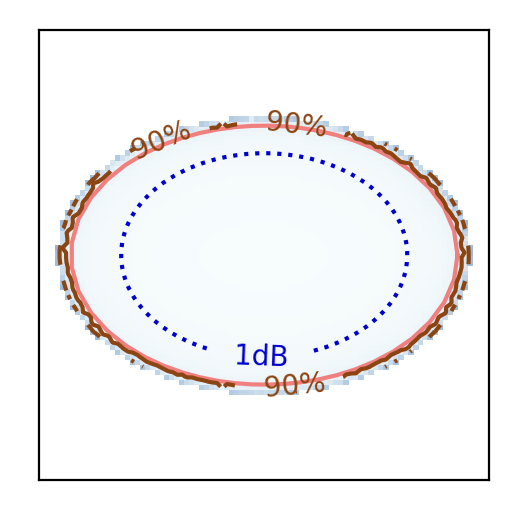}}
 \quad\includegraphics[height=38mm,trim=0mm 5mm 0mm 5mm,clip]{colorbar.png}
 
	\caption{Diffuseness $\psi$ (colormap, solid contour: $90\%$) and level $w$ normalized to origin (dotted \SI{1}{dB} contour) for 2D, 3:2 ellipse of 100 equi-angle uncorrelated vertical line sources (a-d) or 6:4:3 ellipsoid of 2500 maximum determinant nodes with uncorrelated point sources (e-h):
    (a,e) with unity gain $\sigma=1$, 
    (b,f) isotropy-enforcing gain $\sigma=\sqrt{R}$ or $R$, (c,g) gain $\sigma=R$ or $\sqrt{R}^3$, and (d,h) mode-matched $\sigma$ for ideal diffuseness.
	\label{fig:simu4}}
 \vspace{-2mm}
\end{figure*}

This proves with $\alpha=\beta$ that constant sound energy density $\rho\,c^2\,w=\text{const}$ of $2\alpha=D-2$ so $2\beta=D-2$ requires $G_{D}\propto\frac{1}{\sqrt{r}^{D-2}}$; or trivially $G_{D}\propto \frac{1}{r^0}$ of $\beta=0$.

By contrast, this proves with $\alpha=\beta-\frac{1}{2}$ that constant sound-intensity potential $U_\mathrm{I}=\text{const}$  of $2\alpha=D-2$, so $2\beta=D-1$, requires $G_{D}\propto\frac{1}{\sqrt{r}^{D-1}}$, as illustrated before.

\begin{figure}[t]
    \centering
\includegraphics[height=4cm]{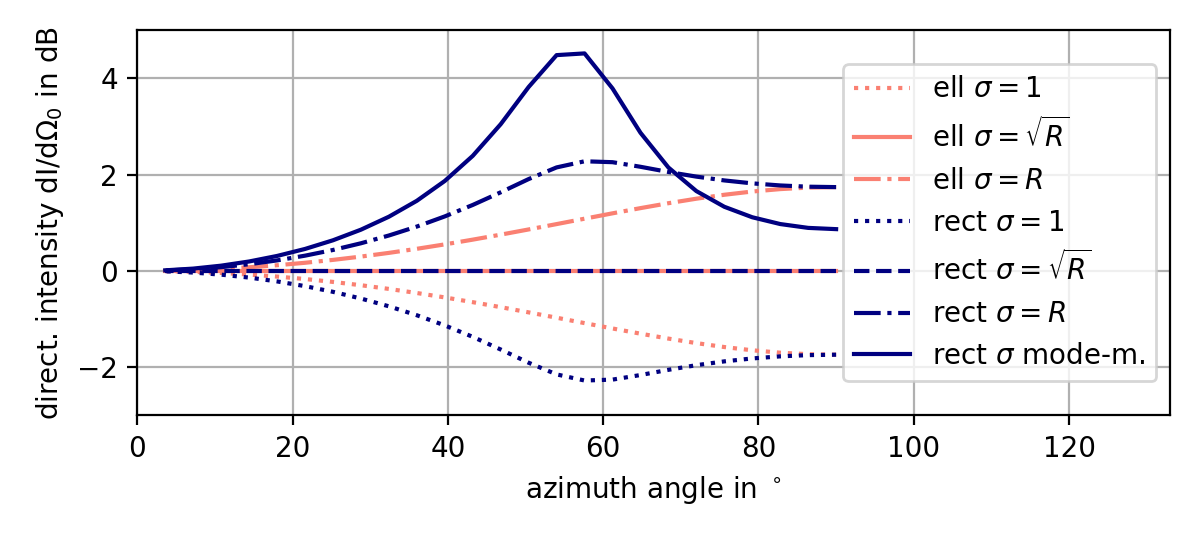}    \caption{Directional intensity $\mathrm{d}I/\mathrm{d}\Omega=\sigma^2/R$ in 2D, normalized at $0^\circ$, observed at $\bm x=\bm 0$ for different azimuth angles to between $0^\circ$ and $90^\circ$ (quadrants are symmetric) for 2:3 2D horizontal ellipse/rectangle of uncorrelated vertical line sources (superellipse $\mathsf{p}=10$) with different gain patterns $\sigma$; isotropy requires $\sigma/\sqrt{R}=\text{const}.$}
    \label{fig:isotropy}
\end{figure}

\begin{figure*}[t]
	\centering
    	\subfigure[superellipse $\sigma=1$]{
 \includegraphics[height=37mm,trim=4mm 4mm 3mm 3mm,clip]{rect_L100_vert_line_srcs_nocomp.png}}	
 \subfigure[superellipse $\sigma=\sqrt{R}$]{
 \includegraphics[height=37mm,trim=4mm 4mm 3mm 3mm,clip]{rect_L100_vert_line_srcs_comp.png}}
  \subfigure[superellipse $\sigma=R$]{
 \includegraphics[height=37mm,trim=4mm 4mm 3mm 3mm,clip]{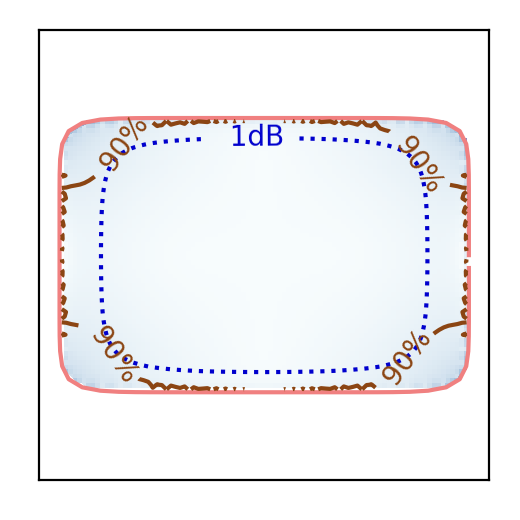}}
   \subfigure[CH mode-match $\sigma$]{
 \includegraphics[height=37mm,trim=4mm 4mm 3mm 3mm,clip]{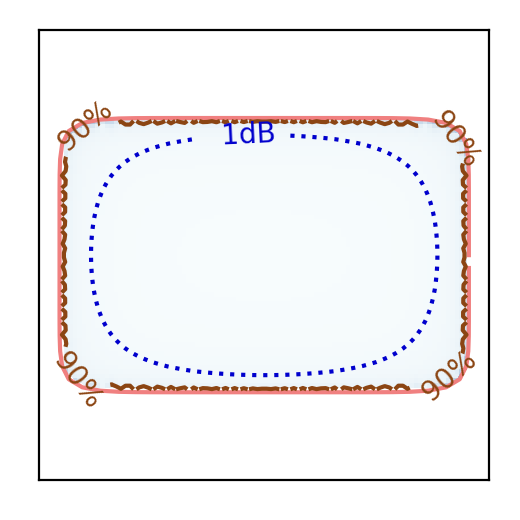}}
 \quad\includegraphics[height=38mm,trim=0mm 5mm 0mm 5mm,clip]{colorbar.png}
 
   	\subfigure[s.ellipsoid/cut1 $\sigma=1$]{
 \includegraphics[height=37mm,trim=4mm 4mm 3mm 3mm,clip]{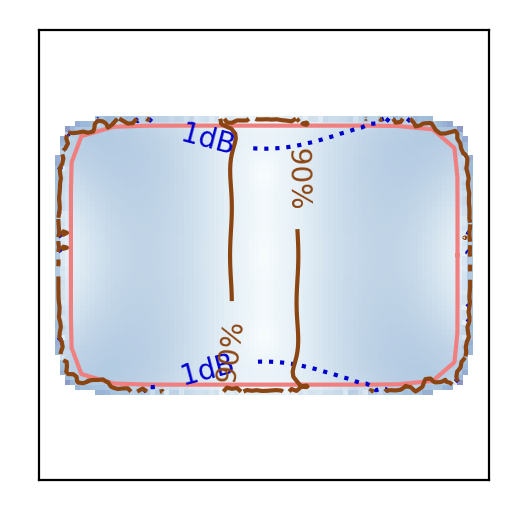}}	
 \subfigure[s.ellipsoid/cut1 $\sigma={R}$]{
 \includegraphics[height=37mm,trim=4mm 4mm 3mm 3mm,clip]{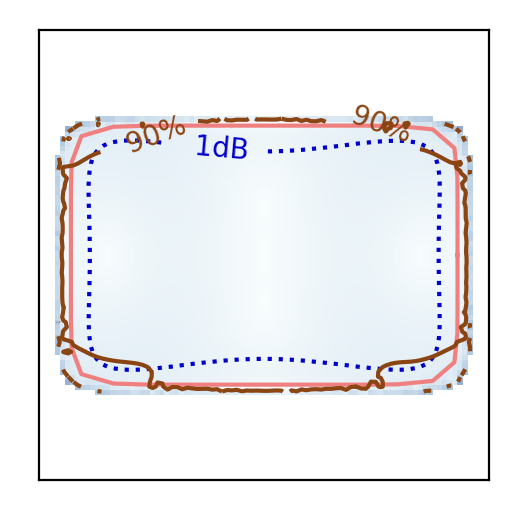}}
  \subfigure[s.ellipsoid/cut1 $\sigma=\sqrt{R}^3$]{
 \includegraphics[height=37mm,trim=4mm 4mm 3mm 3mm,clip]{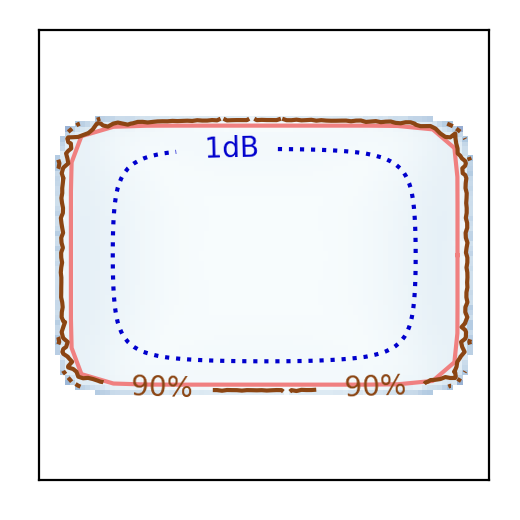}}
   \subfigure[SH mode-match $\sigma$]{
 \includegraphics[height=37mm,trim=4mm 4mm 3mm 3mm,clip]{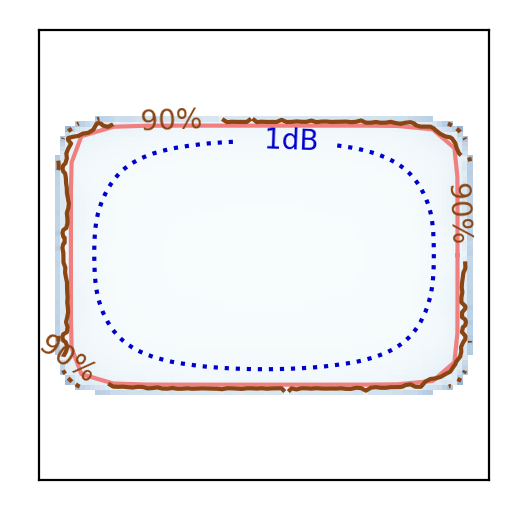}}
  \quad\includegraphics[height=38mm,trim=0mm 5mm 0mm 5mm,clip]{colorbar.png}
    	\subfigure[s.ellipsoid/cut2 $\sigma=1$]{
 \includegraphics[height=37mm,trim=4mm 4mm 3mm 3mm,clip]{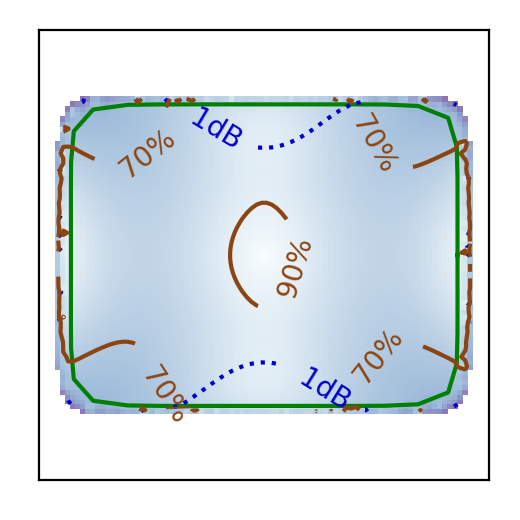}}	
 \subfigure[s.ellipsoid/cut2 $\sigma={R}$]{
 \includegraphics[height=37mm,trim=4mm 4mm 3mm 3mm,clip]{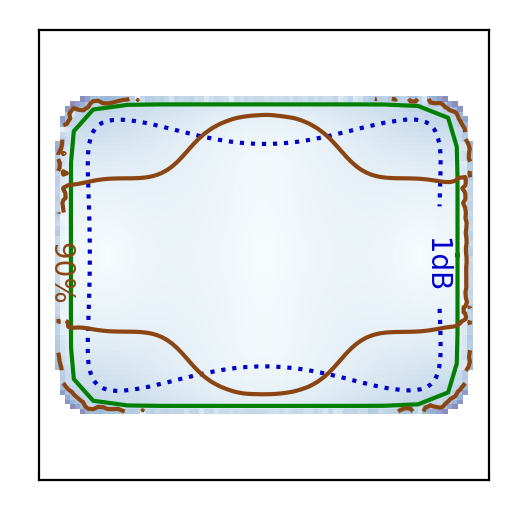}}
  \subfigure[s.ellipsoid/cut2 $\sigma=\sqrt{R}^3$]{
 \includegraphics[height=37mm,trim=4mm 4mm 3mm 3mm,clip]{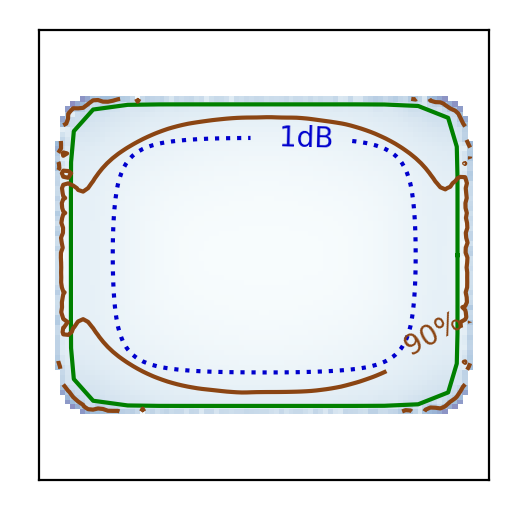}}
   \subfigure[SH mode-match $\sigma$]{
 \includegraphics[height=37mm,trim=4mm 4mm 3mm 3mm,clip]{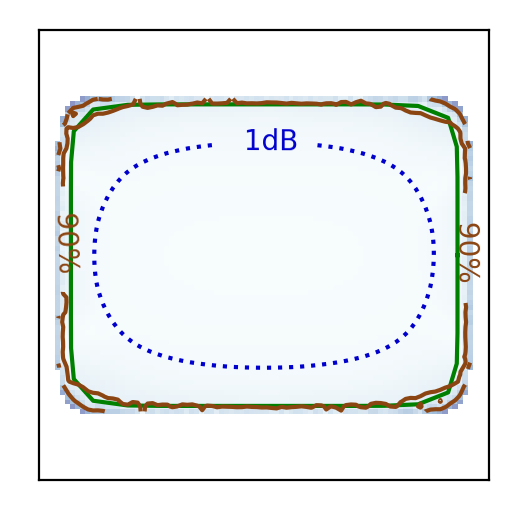}}
 \quad\includegraphics[height=38mm,trim=0mm 5mm 0mm 5mm,clip]{colorbar.png}
 \vspace{-2mm}
	\caption{Diffuseness $\psi$ (colormap, solid contour: $90\%$) and level $w$ normalized to origin (dotted \SI{1}{dB} contour), for the superellipse of the ratio 3:2 (2D), and for the superellipsoid of the ratio  6:4:3 (3D), both with $\mathsf{p}=10$, using 100 equi-angular nodes as uncorrelated line vertical sources (2D) or 2500 maximum determinant nodes as uncorrelated point sources (3D), respectively:     (e,i) with unity gain $\sigma=1$, 
    (f,j) isotropy-enforcing gain $\sigma=R$, (g,k) gain $\sigma=\sqrt{R}^3$, and (h,l) mode-matched $\sigma$ for ideal diffuseness. (i-l) are as (e-h) but plotted along an inclined plane cf.~\autoref{fig:crosssections}.
	\label{fig:simu5}}
 \vspace{-2mm}
\end{figure*}

\section{Suboptimal source layers}\label{sec:suboptimal}
The previous section discussed the shift and rotation invariance of optimal source layers. This section discusses other geometries (ellipsoids, cuboids, incomplete hull).

\subsection{Ellipsoidal source layer}
~\label{sec:ellipsoid}
Ellipsoidal shapes for $D\geq2$ do not provide the above-mentioned symmetry. Their similarity to spherical symmetry bears a chance that source variances can be designed to provide $100\%$ diffuseness. A multi-axial ellipsoid with semi-axes lengths $\bm a=[a_i]$ aligned with the Cartesian axes is described by $\sqrt{\frac{x_\mathrm{1,s}^2}{a_1^2}+\frac{x_\mathrm{2,s}^2}{a_2^2}+\dots}=1$, 
\begin{align}
    \sqrt{\bm x_\mathrm{s}^\intercal\,\mathrm{diag}\{\bm a\}^{-2}\bm x_\mathrm{s}}=1.\label{eq:ellipsoid}
\end{align}
Using a normalized direction vector $\bm u_0$ from the origin, $\|\bm u_0\|=1$, we may write the coordinates as $\bm x_\mathrm{s}=R\,\bm u_0$ to get the direction-dependent radius via eq.~\eqref{eq:ellipsoid},
\begin{align}
   R=\frac{1}{\sqrt{\bm u_0^\intercal\,\mathrm{diag}\{\bm a\}^{-2}\bm u_0}}.\label{eq:ellipsoid_radius}
\end{align}
The distance between source $\bm x_\mathrm{s}$ and receiver at $\bm x=x\,\bm u$, displaced from $\bm 0$ into the direction $\bm u$ by $x$, becomes
\begin{align}
    r&=\sqrt{R^2-2\,R\,x\,\bm u_0^\intercal\bm u+x^2}\nonumber\\
    &=R\,\sqrt{1-2\,\left(\frac{x}{R}\right)\,\bm u_0^\intercal\bm u+\left(\frac{x}{R}\right)^2} \nonumber
\end{align}
Correspondingly, expansion of $\mathcal{G}(r)\propto\frac{1}{r^{2\beta-1}}=\frac{1}{r^{D-2}}$ that was physical and ideal for $U_\mathrm{I}$ above into Gegenbauer polynomials yields 
\begin{align}
    \frac{1}{r^{D-2}}&=\frac{1}{R^{D-2}}\sum_{n=0}^\infty \left(\frac{x}{R}\right)^nC_n^{(\frac{D-2}{2})}(\bm u_0^\intercal\bm u)\\
    &=\sum_{n=0}^\infty x^n\,R^{-(n+D-2)} \, C_n^{(\frac{D-2}{2})}(\bm u_0^\intercal\bm u).\nonumber
\end{align}
We test integrating all sources weighted by the variance 
\begin{align} 
   \sigma^2=R^{D-2+\mu},\label{eq:RDmu}
\end{align}
by integrating over the element $\mathrm{d}\Omega_0=\mathrm{d}\Omega_{0\setminus z}\varpi_\frac{D-2}{2}\mathrm{d}z$ seen from the origin, with $\varpi_\frac{D-2}{2}=\sqrt{1-z^2}^{D-3}$. The idea is to separate the integral over the direction cosine $z=\bm u_0^\intercal\bm u$ to the axis $\bm u$ from other angles $\Omega_{0\setminus z}$ to get
\begin{align}
    U_\mathrm{I}&\propto \sum_{n=0}^\infty x^n\,\int_{-1}^1 \overline{R^{q}}\; C_{n}^{(\frac{D-2}{2})}(z)\,\varpi_{\frac{D-2}{2}}\mathrm{d}z,\label{eq:SigmaRDecomp}\\
    \text{with }\overline{R^{q}}&=\int_{S\setminus z} R^{-(n-\mu)}\;\mathrm{d}\Omega_{0\setminus z}, \nonumber\\
    \text{and } -q&=n-\mu.\nonumber
\end{align}
The integral $\overline{R^{q}}$ is an even function in $z$ as also $R$ is symmetric with regard to a flipped axis $R(\bm u_0)=R(-\bm u_0)$, cf.~eq.~\eqref{eq:ellipsoid_radius}. Consequently, odd $n$ do not contribute, we re-write $n=2l$ and $-q=2l-\mu$, as well as
\begin{align}
    U_\mathrm{I}&\propto\sum_{l=0}^\infty x^{2l}\,\int_{-1}^1 \overline{R^{q}}\;C_{2l}^{(\frac{D-2}{2})}(z)\,\varpi_{\frac{D-2}{2}}\,\mathrm{d}z.
\end{align}
$U_\mathrm{I}=\text{const}$ requires $\int_{-1}^{1} \overline{R^{q}}\,C_{2l}^{(\nu)}\,\varpi_{\nu}\,\mathrm{d}z=\kappa\,\delta_l$
to avoid any degree $l>0$ of variation  $x^{2l}\neq1$.
Integers $-\frac{q}{2}=k\geq0$ and $\frac{\mu}{2}=m$ make $\overline{R^{q}}$  a degree $2k$ polynomial $\mathscr{P}_{2k}$, $\overline{(\bm u_0^\intercal\mathrm{diag}\{\bm a\}^{-2}\bm u_0)^k}$, cf.\  eq.~\eqref{eq:ellipsoid_radius}, or  $\sum_{j=0}^{k}a_{2j}\,C_{2j}^{(\nu)}(z)$ . Orthogonality $\int_{-1}^{1} C_{n}^{(\nu)}\,\,C_{m}^{(\nu)}\,\varpi_{\nu}\,\mathrm{d}z=N_n^2\,\delta_{nm}$  ensures  $\int_{-1}^{1}\mathscr{P}_{2k}\,C_{2l}^{(\nu)}\,\varpi_{\nu}\,\mathrm{d}z=0$ for $l>0$ within $0\leq 2k<2l$. With $2k=2l-2m$, these limits are
$2l\geq 2m>0$ and get most rerstrictive for $l=1>0$, only allowing $m=1$,
\begin{align}
    \mu&=2m=2, &\Rightarrow \sigma&=\sqrt{R}^{D}.
\end{align}
This gain pattern $\sigma$ perfectly accomplishes a constant sound intensity potential $U_\mathrm{I}=\text{const}$ for ellipsoids. 

By contrast, the radial decay of the physical Green's function of the Helmholtz equation is proportional to $|G|\propto\nicefrac{1}{\sqrt{r}^{D-1}}$, cf.~\eqref{eq:green}, 
therefore \emph{isotropy}~\cite{nolan} at $\bm x=\bm 0$ refers to constant directional sound intensity $\frac{\mathrm{d}I}{\mathrm{d}\Omega_0}$, see \autoref{sec:isotropy}. For layers of constant angular density $\mathrm{d}\Omega_0$ instead of constant surface density $\mathrm{d}S$, the term $\frac{\mathrm{d}I}{\mathrm{d}\Omega_0}$ does not contain $\frac{1}{\cos\phi}$ anymore. Only the radiation decay $\frac{1}{\sqrt{R}^{D-2}}$ and source variance $\sigma^2$ remain, so that the intensity arriving from $\mathrm{d}\Omega_0$ around $\bm u_0$ becomes
\begin{align}
    \frac{\mathrm{d}I}{\mathrm{d}\Omega_0}=\frac{\sigma^2}{{R}^{D-1}}.\label{eq:isotropy}
\end{align}
This reveals that sotropy can be enforced by setting the levels $\sigma$ of the surrounding sources to
\begin{align}
    \sigma=\sqrt{R}^{D-1}.
\end{align}

For the 2D case ($D=2$), \autoref{fig:simu4} analyzes the diffuseness and sound energy density of a 3:2 horizontal ellipse of vertical line sources (a-c). Obviously, the unity gain $\sigma=1$ in (a) is not ideal for extended diffuseness greater than $90\%$; yet the diffuseness level already reaches a high level with $>80\%$ everywhere inside.
The isotropy gain $\sigma=\sqrt{R}^{D-1}$ in (b) greatly improves the extent and appears useful, and $\sigma=\sqrt{R}^D$ in (c) readily accomplishes ideal diffuseness everywhere inside, and the equivalent results are accomplished by the mode-matching method (d) presented further below; the sound energy density levels $w$ shown in the figure remain largely unaffected.

\autoref{fig:isotropy} analyzes isotropy by the directional contributions observed at $\bm 0$. For the ellipse of the ratio $a_1:a_2=2:3$, the distant sources are under-represented by the factor $\sqrt{\nicefrac{2}{3}}$, i.e.\ \SI{-1.76}{dB}, with unity gain, see \emph{ell $\sigma=1$} curve. The ideal isotropy weight $\sigma=\sqrt{R}^{D-1}$ removes this shortcoming in the \emph{ell $\sigma=\sqrt{R}$} curve that is consistently flat, while the ideal diffuseness gain $\sigma=\sqrt{R}^D$ is \emph{not isotropic} for the elliptical geometry, see \emph{ell $\sigma=R$} curve. It over-emphasizes distant sources and exaggerates their level by the factor $\sqrt{\nicefrac{3}{2}}$, i.e.\ \SI{1.76}{dB}.

For the 3D case ($D=3$), \autoref{fig:simu4} shows in (e-g) a similar analysis for an ellipsoid with the axis ratio 6:4:3. The cases in (e-g) are convincing already from (f) on and perfect with (g) and (h).  
The sound energy density level $w$ would be nearly perfectly flat for unity gain $\sigma=1$ in (e), but these cases are not so much interesting for their diffuseness: only values slightly higher than $70\%$ are reached in (e) off center. In the other cases, a bit more than a
\SI{1}{dB} increase towards the boundaries has to be accepted.

\subsection{Towards rectangular cuboids:\\ $L_\mathsf{p}$-norm ellipsoid/superellipsoid}\label{sec:superellipse}
To find out if more rectangular cuboid layouts could work similarly, we re-define the multi-axial ellipsoid as superellipsoid~\cite{Barr} with its manifold defined as weighted $L_\mathsf{p}$-norm reaching unity, with semi-axes lengths $\bm a=[a_i]$ as before,
\begin{align}
    \left[\frac{|x_\mathrm{1,s}|^\mathsf{p}}{|a_1|^\mathsf{p}}
+\frac{|x_\mathrm{2,s}|^\mathsf{p}}{|a_2|^\mathsf{p}}+\dots\right]^\frac{1}{\mathsf{p}}&=1.\label{eq:ellipsoidp}
\end{align}
\begin{figure}[t]
\centering
\includegraphics[width=7cm]{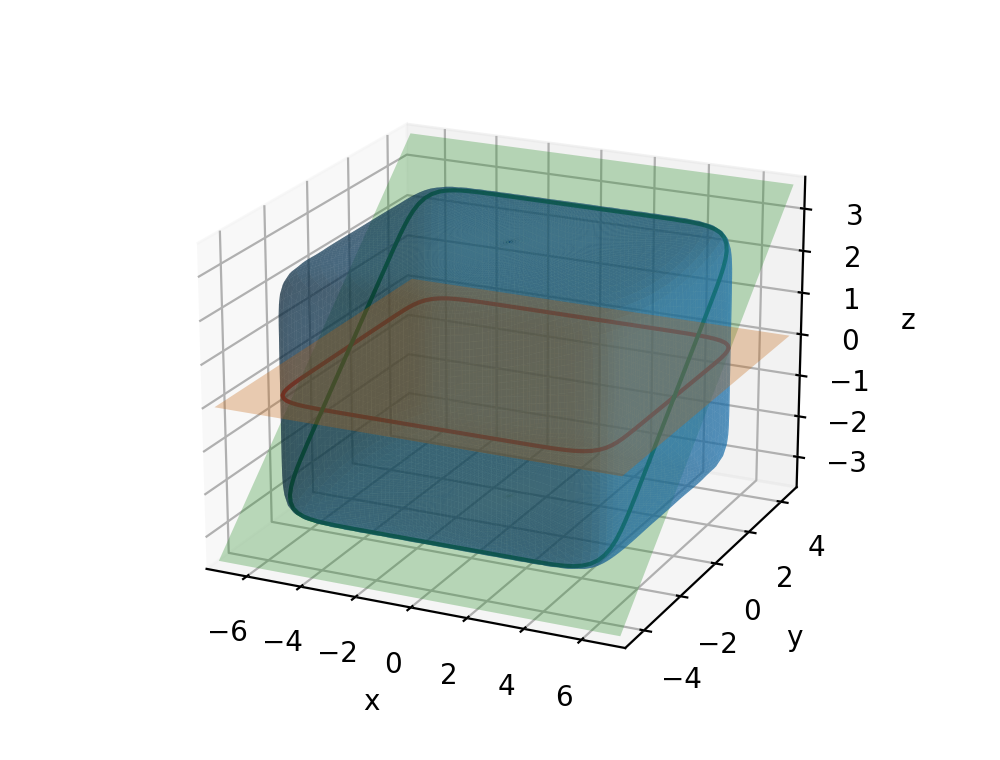}
\caption{Cross-section: 6:4:3 rectangular cuboid (superellipse $\mathsf{p}=10$): cut rotated by $0^\circ$ or $36.9^\circ$ wrt.\ $x$.\label{fig:crosssections}\vspace{-2mm}
}
\end{figure}
The higher $2<\mathsf{p}\rightarrow\infty$, the more cuboid the surface gets.
With the entries $u_{0,i}$ of the direction vector $\bm u_0$, we  write the coordinates as $ x_{\mathrm{s},i}=R\, u_{0,i}$ and get 
\begin{align}
   R=\frac{1}{\left(
   \sum_{i=1}^D \frac{|u_\mathrm{0,i}|^\mathsf{p}}{|a_i|^\mathsf{p}}\right)^\frac{1}{\mathsf{p}}}.\label{eq:ellipsoidp_radius}
\end{align}
And this effectively shows why it is not as easy as before to find an ideal weight $\sigma$ as for the ellipsoid: Here, ${R^{\mu-n}}= \left(\sum_{i=1}^D \frac{|u_\mathrm{0,i}|^\mathsf{p}}{|a_i|^\mathsf{p}}\right)^\frac{n-\mu}{\mathsf{p}}$ becomes a $\mathsf{p}k$-degree polynomial with integer $k$ and $n-\mu=\mathsf{p}k>0$ to remove the root, and even $\mathsf{p}>0$ to avoid the absolute value. As $\overline{R^{\mu-n}}$ is even, $n=2l$ and $2l-\mu=\mathsf{p}k$ must hold for integer $l>0$ and constant $\mu$, we get $\mathsf{p}=2$ and $\mu=\mathsf{p}m$. This only leaves the ellipsoid case discussed in \autoref{sec:ellipsoid} to provide a simple solution for diffuseness with $\sigma=\sqrt{R}^D$. 

For the 2D case ($D=2$), \autoref{fig:simu5} analyzes in (a-c) the diffuseness accomplished with the directional gains $\sigma=1$, $\sigma=\sqrt{R}^{D-1}$, $\sigma=\sqrt{R}^{D}$ for a 3:2 horizontal superellipse with $\mathsf{p}=10$ and uncorrelated vertical line sources. In fact, the drop of diffuseness in the corners underlines that there is no simple ideal solution. And yet, ideal isotropy gains $\sigma=\sqrt{R}^{D-1}$ in (b) produce diffuseness inside that already exceeds $70\%$; the sound energy density levels $w$ shown in the figure remain largely unaffected (a-c).
The isotropy gain $\sigma=\sqrt{R}^{D-1}$ in (b) greatly improves the extent, slightly more does $\sigma=\sqrt{R}^D$ in (c), but only the mode-matching solution (d) presented below also reaches the corners; the sound energy density levels $w$ behave similarly for (e-h) as for the elliptic cases (a-d).

\autoref{fig:isotropy} also analyzes isotropy by the corresponding directional contributions in the superellipse layout, see \emph{rect} curves. The noticable difference to the \emph{ell} curves of the ellipse is around $56^\circ$, the angle targeting the corner of the circumscribed 3:2 rectangle, at which the maximum radius $\max R$ is reached. There, $\sigma=1$   yields the minimum directional contribution, and $\sigma=R$ a maximum one, while $\sigma=\sqrt{R}$ is the ideal isotropy solution $\sigma=\sqrt{R}^{D-1}$ for $D=2$.

The 3D ($D=3$) cuboid case is analyzed in \autoref{fig:simu5}(e-l)  for the superellipsoid with the axes ratio 6:4:3. 
The $xy$ cross-section (e-h) only cuts the vertical edges, which seem to be uncritical compared to the 2D case. Therefore, most of the diagrams already look convincing (f-h). For instance the levels $\sigma=\sqrt{R}^{D}$ in (g) appear to be a perfect choice, and mode-matching in (h) does not yield further  improvement on the $xy$ cut.
However the inclined cross-section (i-l), cf.~\autoref{fig:crosssections}, cuts the corners of the cuboid and reveals that full coverage with $90\%$ diffuseness is also problematic in the cases (i-k).

\begin{figure*}
    \centering
    \subfigure[ellipsoid $\sigma=1$]{\includegraphics[height=3.7cm,trim=6mm 0mm 28mm 0mm,clip]{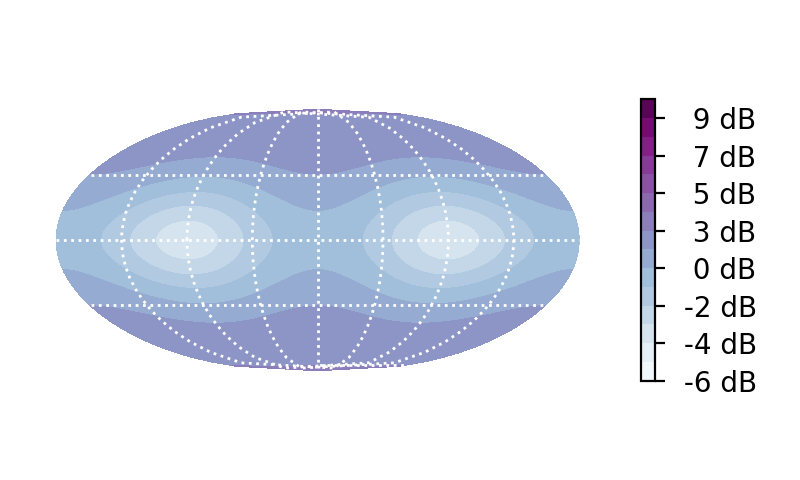}}
    \subfigure[ellipsoid $\sigma=R$]{\includegraphics[height=3.7cm,trim=6mm 0mm 28mm 0mm,clip]{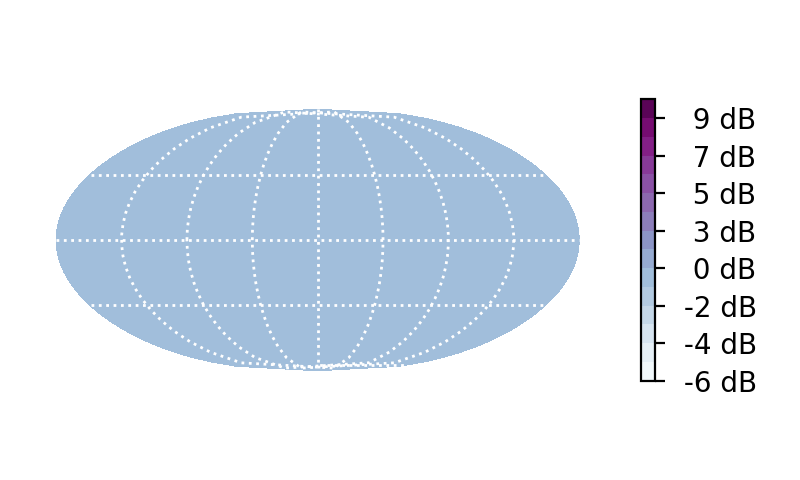}}
    \subfigure[ellipsoid $\sigma=\sqrt{R}^3$]{\includegraphics[height=3.7cm,trim=6mm 0mm 28mm 0mm,clip]{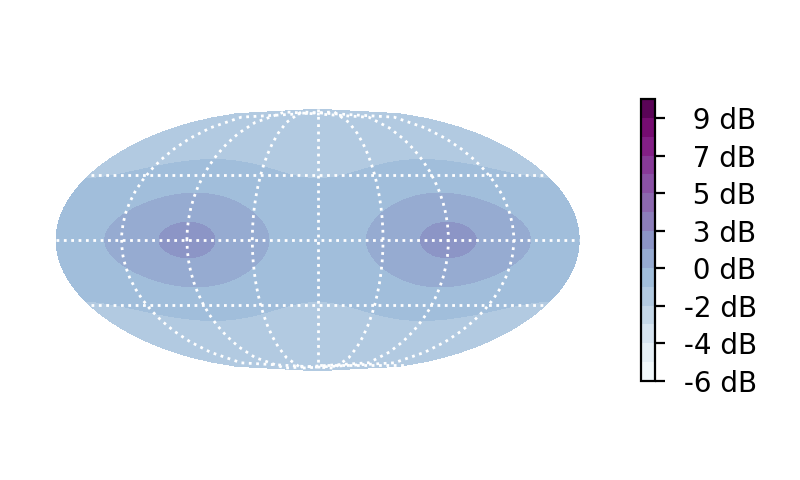}}
    \subfigure[ellipsoid $\sigma$ modematch]{\includegraphics[height=3.7cm,trim=6mm 0mm 28mm 0mm,clip]{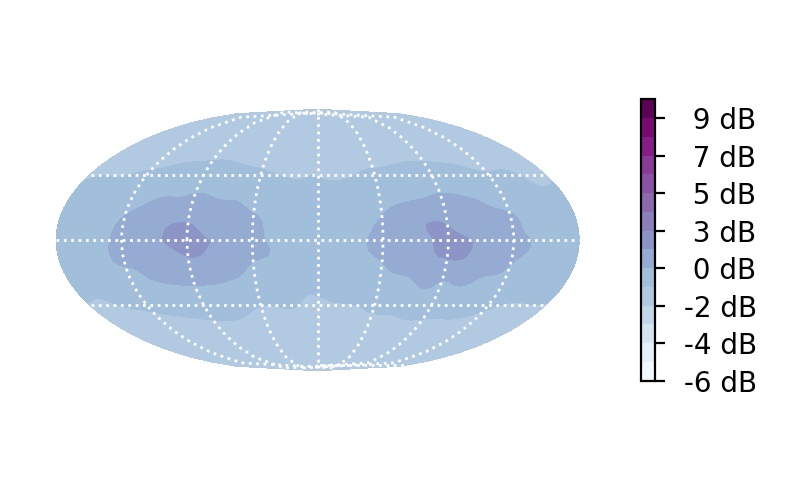}}
    \includegraphics[height=3.7cm,trim=8cm 0mm 5mm 0mm,clip]{3d_isotropy_ell3g1.png}
        \subfigure[cuboid $\sigma=1$]{\includegraphics[height=3.7cm,trim=6mm 0mm 28mm 0mm,clip]{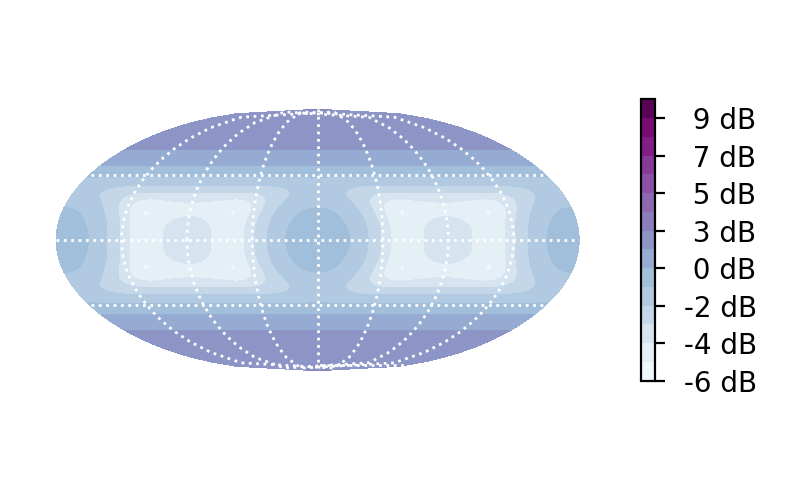}}
    \subfigure[cuboid $\sigma=R$]{\includegraphics[height=3.7cm,trim=6mm 0mm 28mm 0mm,clip]{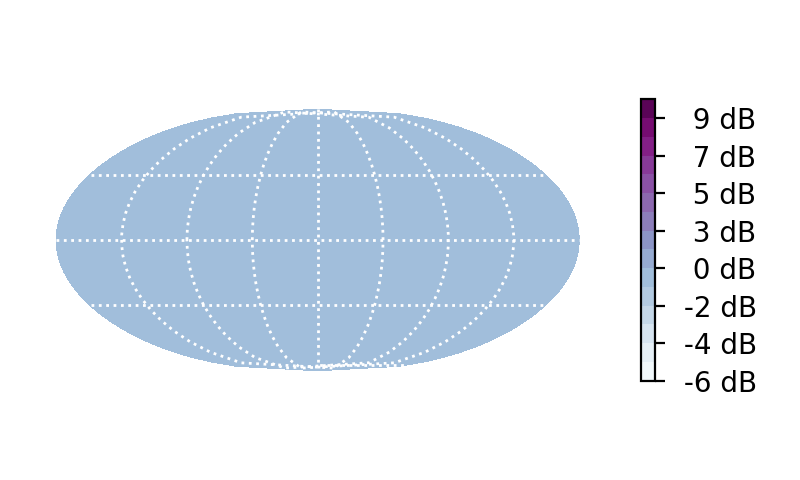}}
    \subfigure[cuboid $\sigma=\sqrt{R}^3$]{\includegraphics[height=3.7cm,trim=6mm 0mm 28mm 0mm,clip]{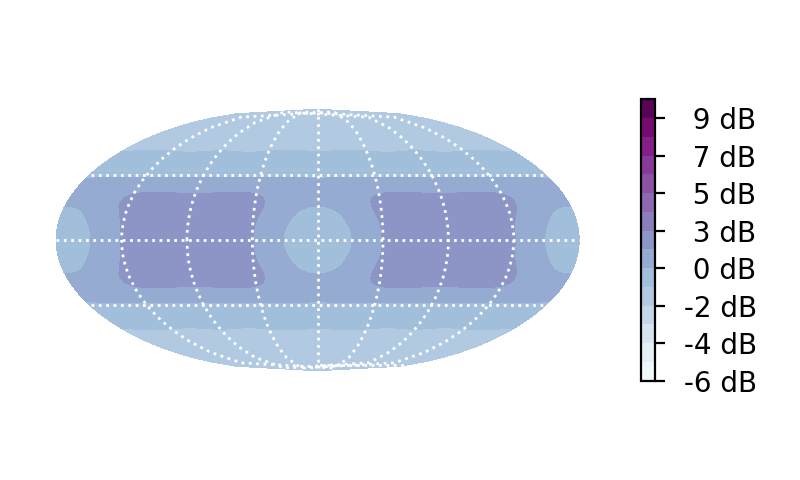}}
    \subfigure[cuboid $\sigma$ modematch]{\includegraphics[height=3.7cm,trim=6mm 0mm 28mm 0mm,clip]{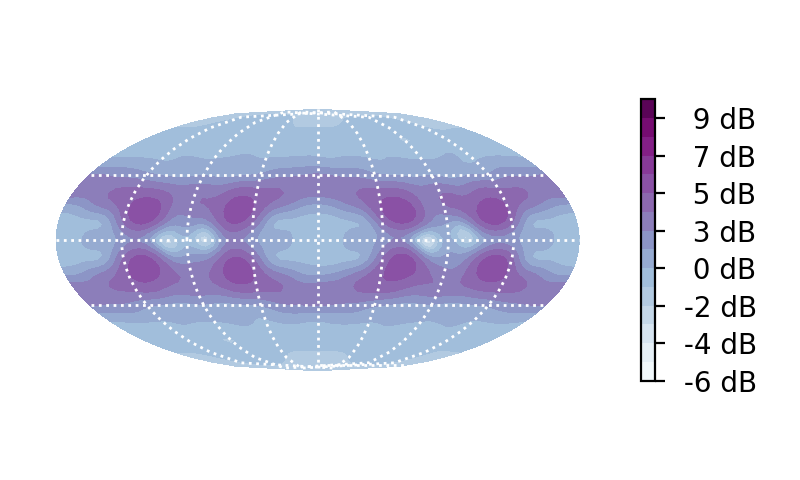}}
    \includegraphics[height=3.7cm,trim=8cm 0mm 5mm 0mm,clip]{3d_isotropy_rec3g1.png}
    \caption{Directional intensity $\mathrm{d}I/\mathrm{d}\Omega=\sigma^2/R^2$ in 3D mapped as $(\varphi\cdot|\sin\vartheta|^{0.4}$, $\frac{\pi}{2}-\vartheta)$ on $\varphi\in[-\pi,\pi]$,  $\vartheta\in[0,\pi]$, normalized at $(\varphi,\vartheta)=(0,\frac{\pi}{2})$, for $\bm x=\bm 0$, 
    cf.~eq.~\eqref{eq:isotropy}, with different choices of directional magnitude $\sigma=\{1,R,R^{1.5}\}$ or mode matched, for a 6:4:3 ellipse (top) or superellipsoid cuboid (bottom) in 3D, i.e.\ with $D=3$.\label{fig:isotropy3d}}
\end{figure*}

\autoref{fig:isotropy3d}  analyzes the isotropy for the ellipsoid in (a-d) by the directional contributions for different choices of $\sigma$. The isotropy-enforcing levels $\sigma=R$ accomplish perfectly flat results, while for $\sigma=1$ the most distant direction is quieter by about $-\SI{6}{dB}$, and in (c) $\sigma=\sqrt{R}^3$, it is louder by $\SI{4}{dB}$.

If the resulting diffuseness is insufficient for a given, arbitrary non-circular or non-spherical layout, despite the promising examples in  \autoref{fig:simu5}(f) or (g). Only mode-matching (l) proposed in \autoref{sec:modematch2} and  \autoref{sec:modematch3} enforces ideal diffuseness reaching the corners.

\subsection{Mode matching: non-circular}
\label{sec:modematch2}
For $D=2$, Green's function $\mathcal{G}=-\frac{\ln r}{2\pi}$ of the potential is with $r=R\sqrt{1-2z\rho+\rho^2}$, $z=\cos(\varphi-\varphi_0)$, $\rho=\frac{x}{R}$,
\begin{align}
   -\mathcal{G}&=\frac{\ln r}{2\pi}=\lim_{\nu\rightarrow0}\sum_{n=1}^\infty \rho^n \frac{C_{n}^{(\nu)}(z)}{2\pi}=2\sum_{m=1}^\infty \rho^m \frac{T_{m}(z)}{2\pi\,m}\nonumber\\
   &=\sum_{m=-\infty}^\infty \frac{x^{|m|}}{R^{|m|}} \frac{(1-\delta_m)\,\Phi_m(\varphi)\Phi_m(\varphi_0)}{|m|},
\end{align}
for a source at polar coordinates $R,\varphi_0$ evaluated at $x,\varphi$,
with Chebyshev polynomials $T_m(z)$ and circular harmonics
\begin{align}
   \Phi_m(\varphi)&=\frac{1}{\sqrt{2\pi}}
   \begin{cases}
   \sqrt{2}\sin(m\varphi),&m<0\\
   1, &m=0\\
   \sqrt{2}\cos(m\varphi),&m>0.
   \end{cases}
\end{align}
Superimposing multiple uncorrelated sources with regard to the angle $\varphi_0$ via an integral, their angle-dependent radius $R$ and variance $\sigma^2$, we can represent the total contribution by their coefficient $a_m$
\begin{align}
    U_\mathrm{I}&=2\sum_{m=-\infty}^\infty \rho^m \,\Phi_m(\varphi)\,(1-\delta_m)\,a_m\\
    a_m&=
    \int_{0}^{2\pi}\frac{\Phi_m(\varphi_0)}{|m|\,R(\varphi_0)^{|m|}}\;\sigma(\varphi_0)^2\;\mathrm{d}\varphi_0.\nonumber
\end{align}
For an ideally diffuse intensity potential $U_\mathrm{I}=0=\text{const}$, we try to find variances $\sigma^2$ providing $a_m=0$ for $m>0$.

Numerically, we can implement the search for a limited $|m|\leq N$ with reasonably large maximum degree $N$ to avoid as many $|m|\neq0$ terms getting synthesized as possible. We may accordingly uniformly sample the source angles $\varphi_{0,l}=\frac{2\pi}{L}\,l$ and write the integral for $a_m$ as summation 
\begin{align}
   a_m=\sum_{l=1}^L\frac{\Phi_m(\varphi_{0,l})}{|m|\,R(\varphi_{0,l})^{|m|}}\,\sigma_l^2\;\frac{2\pi}{L}.\nonumber
\end{align}
It can be defined as matrix vector product equation with the regularization $|m|\rightarrow \max\{|m|,1\}$ to find the source variances $\bm \sigma_2=[\sigma_l^2]$
\begin{align}
   \mathbf{i}_0&=\bm M\,\bm \sigma_2,
   &\mathbf{i}_0&=[\delta_m]_m,&\bm \sigma_2&=[\sigma_l^2]_l\nonumber\\
   \bm M&=
\bigg[\frac{\Phi_m(\varphi_{0,l})}{|m|\,R_l^{|m|}}\bigg]_{m}^l,
   \nonumber\\
   \bm \sigma_2&=\bm M^{-1}\mathbf{i}_0.
\end{align}
For regularization, either the expression $R(\varphi_{0,l})^{-|m|}$ should be damped towards $|m|\rightarrow\infty$, or $N$ should be limited at some point. The simulations below still worked fine without additional regularization with $L=100$ and selecting the modes $-49\leq m\leq 50$ to get a $100\times 100$ square matrix $\bm M$.

\autoref{fig:simu4}(d) shows the mode-matching solution for the horizontal 3:2 ellipse of uncorrelated vertical line sources, which is obviously the optimum $\sigma=R$ in (c) obtained analytically, before. For the $\mathsf{p}=10$ superellipse, \autoref{fig:simu5}(d) shows the improvement to ideal diffuseness. 
While this might be helpful, the isotropy of the ideal solution might not always be satisfactory: \autoref{fig:isotropy} shows in its \emph{rect $\sigma$ mode-m.}-curve that the contribution of any corner source is largely over-emphasized. Over emphasis gets even worse for $\mathsf{p}>10$. Nevertheless, on sparse discrete-direction layouts, the behavior may depend on the detail, i.e., how corners are sampled.

\subsection{Mode matching: non-spherical}
\label{sec:modematch3}
For $D=3$, the Green's function $\mathcal{G}=\frac{1}{r}$ of the potential is with $r=R\sqrt{1-2z\rho+\rho^2}$, $z=\cos(\bm u^\intercal\bm u_0)$, $\rho=\frac{x}{R}$,
\begin{align}
   \mathcal{G}&=\frac{1}{4\pi r}=\frac{1}{R}\sum_{n=0}^\infty \rho^n \frac{C_{n}^{(\frac{1}{2})}(z)}{4\pi}=\frac{1}{R}\sum_{n=0}^\infty \rho^n \frac{P_{n}(z)}{4\pi}\nonumber\\
   &=\sum_{n=0}^\infty\sum_{m=-n}^n \frac{x^n}{R^{n+1}} \,\frac{Y_n^m(\bm u)\,Y_n^m(\bm u_0)}{2n+1},
\end{align}
for a source at $R\,\bm u_0$ evaluated at $\bm x=x\,\bm u$, with the spherical harmonics $Y_n^m$ at the azimuth $\varphi=\arctan\frac{u_2}{u_1}$ and zenith $\vartheta=\arccos u_3$ observed in the entries of $\bm u=[u_i]_i$
\begin{align}
Y_n^m(\bm u)&=N_n^m\,\Phi_m(\varphi)\,P_n^{|m|}(\cos\vartheta)\\
   N_n^m&=(-1)^m\sqrt{\frac{2n+1}{2}\frac{(n-|m|)!}{(n+|m|)!}}.\nonumber
\end{align}
Integrating over uncorrelated sources that are uniformly distributed across all directions $\bm u_0$ with their radius $R$ and source variance $\sigma^2$, we get
\begin{align}
    U_\mathrm{I}&=\sum_{n=0}^\infty\sum_{m=-n}^n x^m \,Y_n^m(\bm u)\,a_{nm}\\
    a_{nm}&=\int_{\mathbb{S}^2}\frac{Y_n^m(\bm u_0)\,\sigma(\bm u_0)^2}{(2n+1)\,R(\bm u_0)^{n+1}}\;\mathrm{d}\bm u_0.\nonumber
\end{align}
We can find suitable variances $\sigma_l^2$ that ensure $a_{nm}=0$ for $n>0$ to get a constant intensity potential $U_\mathrm{I}=\text{const.}$ Similar as above, we define the integral for $a_{nm}$ as sum for numeric evaluation
\begin{align}
   a_{nm}&=\sum_{l=1}^L\frac{Y_n^m(\bm u_{0,l})\,\sigma_l^2}{(2n+1)\,R_l^{n+1}}\;\frac{4\pi}{L}.\nonumber
\end{align}
And ideal discretization for numerical integration can be found in spherical $t$-designs, for instance. 
The squared gains $\sigma_l^2$ are found by their modeling in terms of a spherical harmonics expansion with the coefficients $\gamma_n^m$ writing the summation as matrix-vector product
\begin{align}
    \bm \sigma_2&=\bm Y\,\bm \gamma_2,&
   \bm \sigma_2&=[\sigma_l^2]_l,\nonumber\\
    \bm Y&=[Y_n^m(\bm u_{0,l})]_{nm}^l
, 
  &\bm\gamma_2&=[\gamma_n^m]_{nm},\nonumber\\
   \bm M&=
\bigg[\frac{Y_n^m(\bm u_{0,l})}{(2n+1)R_l^{n+1}}\bigg]_{nm}^l.
\end{align}
The procedure to synthesize only a zeroth-order component is different here, as $\bm M$ has a poor numerical conditioning.
Instead, to avoid such coefficients to get synthesized, the $n=0$ row of $\bm M$ is removed $\bm M_{\setminus 0}$ and should vanish when multiplied with solution variances $\bm\sigma_2=\bm Y\,\bm \gamma_2$, 
\begin{align}
   \bm M_{\setminus 0}\,\bm Y\,\bm\gamma_2=\bm 0,
\end{align}
and we may chose to regularize the result by also minimizing higher-order coefficients in $\gamma_n^m$ by a weight $w_n$ in $\bm W=[w_n]_{nm}$
\begin{align}
   \tilde{\bm M}_{\mathrm{W}\setminus0}=\begin{bmatrix}
      \bm M_{\setminus 0}\,\bm Y\\
      \bm W
      \end{bmatrix}\,\bm\gamma_2\rightarrow\min.
\end{align}
In the analysis in \autoref{fig:simu5}(d), (h), (l), the maximum-determinant points~\cite{Womersley} from the website~\cite{WomersleyURL} were chosen for $N=49$ and $L=2500$, with the idea to get a square matrix $\bm M$. However conditioning was $\mathrm{cond}\{\bm M\}>10^{22}$, so regularization is necessary in the spherical case, for which the order was limited to $N=17$, despite $49$ would be possible. For further regularization that avoids unnecessary oscillation in $\bm\sigma_2$, a suitably normalized, wiggle-suppressing weight
\begin{align}
   w_n={\textstyle\frac{0.015}{\sqrt{4\pi\,\sum_n' [n'(n'+1)]^2}}}\;n\,(n+1)
\end{align}
was used, cp.~\cite[p.428,Eq.(18.74)]{boyd},
so that of the SVD $\tilde{\bm M}_{\mathrm{W}\setminus0}=\bm U\mathrm{diag}_{L\times(N+1)^2}\{\bm s\}\bm V^\intercal$ the vector $\bm\gamma_2=\bm v_\mathrm{end}$
of the smallest singular value $s_\mathrm{end}$ could be used as minimizer that yields $\bm\sigma_2=\bm Y\,\bm\gamma_2$.

Same as for the elliptic case (2D) but in the ellipsoidal 3D case, the mode-matching solution in \autoref{fig:simu4}(h) brings no further improvement over the analytic solution $\sigma=\sqrt{R}^{D}$ in (g). An improvement is obtained in the superellipsoidal case, at least when inspecting the differences in the cross sections \autoref{fig:simu5}~(k) and (l) including the corners of the cuboid. Particularly the regularized mode-matching
solution pushes the $90\%$ diffuseness synthesis outwards to the corners of the cuboid, while sightly increasing the level towards them, as seen in the more rounded shape of the $\SI{1}{dB}$ contour for $\rho\,c^2\,w$ in (l).

Isotropy of the mode-matching solution is analyzed in \autoref{fig:isotropy3d}(d) for the ellipsoid case, with not much difference to the analytic $\sigma=\sqrt{R}$.

\subsection{Hemisphere}\label{sec:hemisphere}
As practical realizations of spherical (3D) loudspeaker layouts often only cover the upper hemisphere with point sources $\beta=1$, a numerical simulation was carried out to analyze the resulting sound field metrics.
\begin{figure}[h]
	\centering
 \includegraphics[height=37mm,trim=8mm 7mm 6mm 7mm,clip]{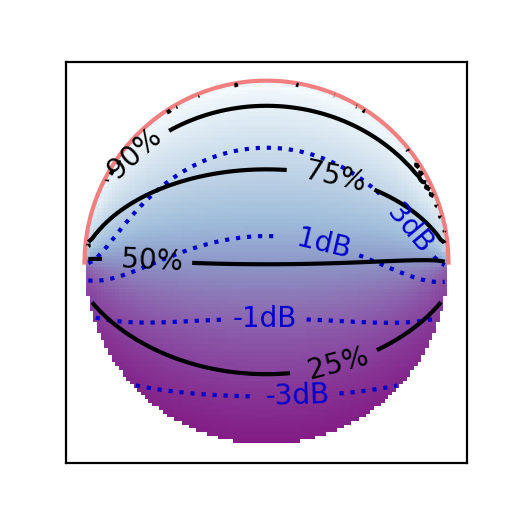}
\quad\includegraphics[height=38mm,trim=0mm 5mm 0mm 5mm,clip]{colorbar.png}
	\caption{Sound energy density (dotted contours) and diffuseness $\psi$ (colors+contours)
	for 2598 of 5200 upper-hemisphere point sources (Gr\"af, Chebyshev-type~\cite{graefurl}).
	\label{fig:hemisphere}}
\end{figure}

\begin{figure*}[t]
	\centering
	\subfigure[circle of 4 vert. line srcs.]{
 \includegraphics[height=37mm,trim=4mm 4mm 3mm 3mm,clip]{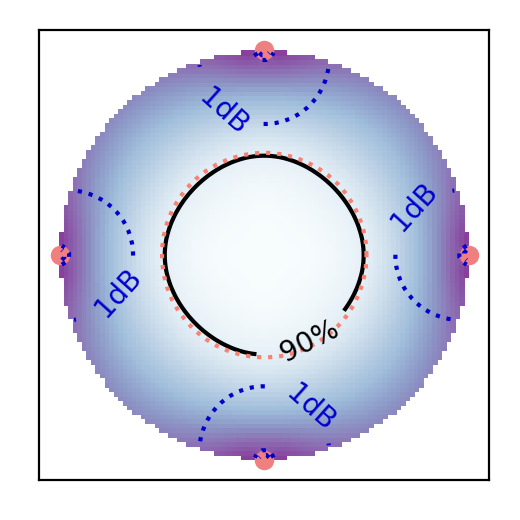}}	\subfigure[circle of 6 vert.\ line srcs.]{
 \includegraphics[height=37mm,trim=4mm 4mm 3mm 3mm,clip]{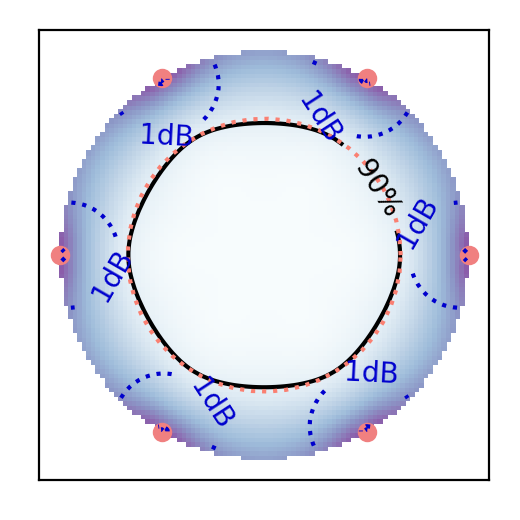}}
     \subfigure[circle of 8 vert.\ line srcs.]{
 \includegraphics[height=37mm,trim=4mm 4mm 3mm 3mm,clip]{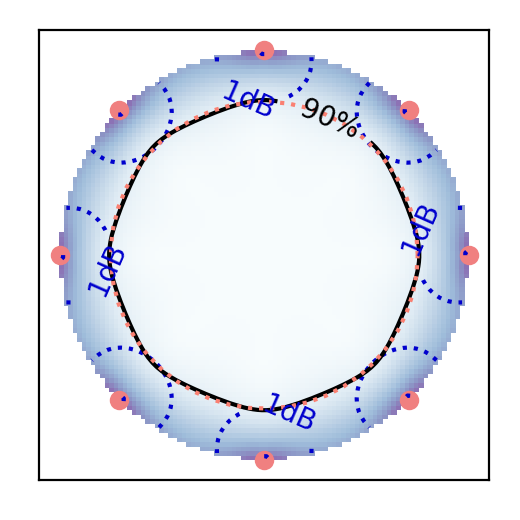}}
     \subfigure[circle of 12 vert.\ line srcs.]{
 \includegraphics[height=37mm,trim=4mm 4mm 3mm 3mm,clip]{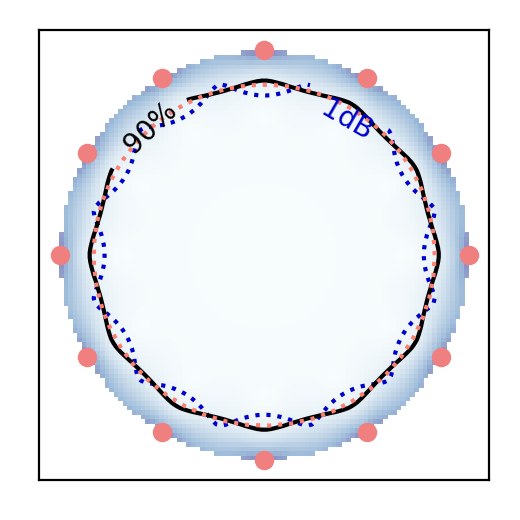}}
\quad\includegraphics[height=38mm,trim=0mm 5mm 0mm 5mm,clip]{colorbar.png}
	\caption{Diffuseness $\psi$ (colormap and $90\%$ contour) and level $w$ normalized to origin (blue dotted \SI{1}{dB} contour) of a unit circle of $L=\{4,6,8,12\}$ equi-angle uncorrelated vertical line sources (dots), compared against a circle of the radius 
    $r=\{\nicefrac{1}{2},
    \nicefrac{2}{3},
    \nicefrac{3}{4},
    \nicefrac{5}{6}\}$ (orange red, dotted).
	\label{fig:simu2}}
\end{figure*}
\begin{figure*}[t]
	\centering
	\subfigure[sphere of 6 pt. srcs.]{
 \includegraphics[height=37mm,trim=4mm 4mm 3mm 3mm,clip]{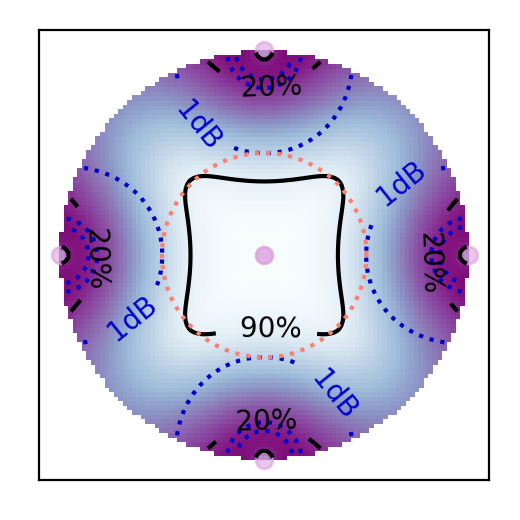}}	\subfigure[sphere of 12 pt. srcs.]{
 \includegraphics[height=37mm,trim=4mm 4mm 3mm 3mm,clip]{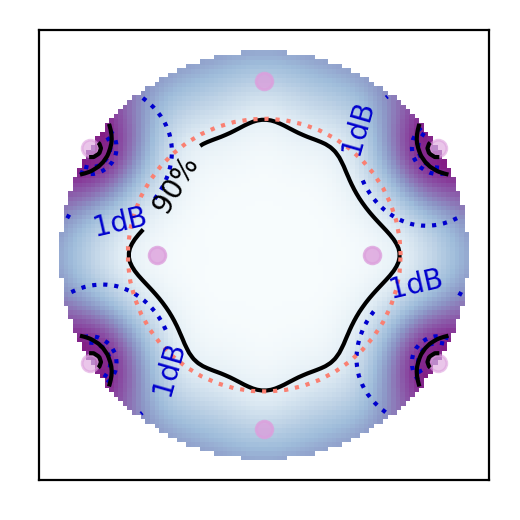}}
     \subfigure[sphere of 24 pt. srcs.]{
 \includegraphics[height=37mm,trim=4mm 4mm 3mm 3mm,clip]{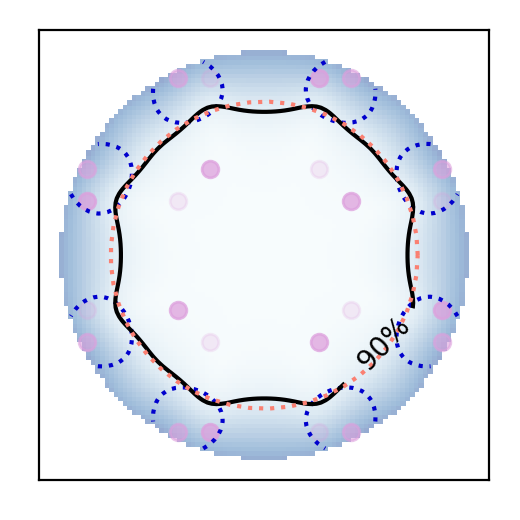}}
     \subfigure[sphere of 70 pt. srcs.]{
 \includegraphics[height=37mm,trim=4mm 4mm 3mm 3mm,clip]{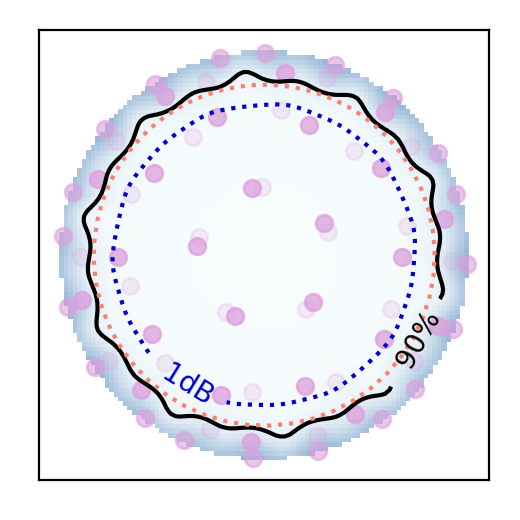}}
\quad\includegraphics[height=38mm,trim=0mm 5mm 0mm 5mm,clip]{colorbar.png}
	\caption{Diffuseness $\psi$ (colormap and $90\%$ contour) and level $w$ normalized to origin (blue dotted \SI{1}{dB} contour) of a unit sphere of $L=\{6,12,24,70\}$ point sources (dots) from spherical $\{3,5,7,11\}$ designs (Gr\"af, Chebyshev-type~\cite{graefurl}) compared against a radius 
    $r=\{\nicefrac{1}{2},
    \nicefrac{2}{3},
    \nicefrac{3}{4},
    \nicefrac{5}{6}\}$ (orange red, dotted).
	\label{fig:simu6}}
\end{figure*}

At the horizontal cutting plane that bounds the upper hemisphere, the missing lower hemisphere causes a loss of up-down symmetry. Where up-down symmetry cancelled the vertical intensity component before, it becomes non-zero $I_\mathrm{z}<0$ along a horizontal observer position at $z=0$.
The hemispherical integrals 
\begin{align}
	\rho\,c^2\,w&=\frac{1}{2\pi}\int_0^1\,\int_0^{2\pi}
	  \frac{\mathrm{d}\cos\vartheta\,\mathrm{d}\varphi}{\|\bm x-\bm u(\varphi,\vartheta)\|^{2}},\\
	  \rho\,c\,\bm I&=\frac{1}{2\pi}\int_0^1\,\int_0^{2\pi}
	  \frac{[\bm x-\bm u(\varphi,\vartheta)]\,\mathrm{d}\cos\vartheta\,\mathrm{d}\varphi}{\|\bm x-\bm u(\varphi,\vartheta)\|^{3}},
\end{align}
are easy to solve at $\bm x=\bm 0$ where the denominator is unity, sound energy density is unity $\rho\,c^2\,w=\int_0^1\mathrm{d}z=1$, horizontal sound intensities vanish because of rotational symmetry, and the vertical one is $\rho\,c\,I_z=-\int_0^1 z\,\mathrm{d}z=-\nicefrac{1}{2}$, yielding a diffuseness of $\psi=50\%$. For \autoref{fig:hemisphere}, the integrals were solved numerically via a spherical $t$-design of high value for $t=100$ with $L=5200$ nodes, of which about $\nicefrac{L}{2}$ lie on the upper hemisphere. Diffuseness stays roughly $50\%$ on the horizontal cut.
At vertical shifts into the hemisphere, diffuseness exceeds $75\%$ at about half  of the height.

\section{Discrete source layouts: $t$-designs}\label{sec:discrete}

The integral over the Gegenbauer expansions of eqs.~\eqref{eq:wpot} \eqref{eq:ipot} on the unit sphere can be ideally discretized and  holds for a maximum degree $N\leq t$ with discrete sourced arranged in a spherical $t$-design~\cite{Delsarte}.
Knowing that only the $n=0$ component is  present around $x\rightarrow0$, and gradually with rising $x$ the next higher expansion powers emerge, the first uncontrolled component  $x^{N+1}C_{N+1}^{(\nu)}(z)$ defines the error. Assuming the RMS value of $C_{N+1}^{(\nu)}(z)$ in $z$ is sufficiently close to unity, we get an approximate radial limit of an error-free sweet area 
\begin{align}
   x^{N+1}&<e^{-1}\nonumber\\
   x&<e^{-\frac{1}{N+1}}\approx 1-{\textstyle\frac{1}{N+1}}.\label{eq:sweet_x}
\end{align}
This implies that the synthesized diffuse field gets larger by increasing the number of spherical $t$-design nodes which reach a higher values for $t\geq N$, see also~\cite{tanaka23}.

\subsection{Discretization study 2D}
\label{sec:discretization_study2d}
Uncorrelated vertical line sources evenly arranged in a circle are the ideal case, and discretized to only $L=\{4,6,8,12\}$ evenly spaced sources, simulation yields the maps in \autoref{fig:simu2}~(a-d). Obviously, the relative sweet area within which diffuseness exceeds $\psi\geq90\%$ is reduced and scales with the number of sources. Scaling is $100\%\cdot\frac{L-2}{L}$ as drawn in orange red. For a radius $r$ in eq.~\eqref{eq:sweet_x} following this contour, $t$ is chosen to be $t=2N+1$ for  the $2N+2$ equi-angle nodes.

\subsection{Discretization study 3D}
\label{sec:discretization_study}
For the sphere, spherical $t$-designs~\cite{Delsarte} are useful discretization schemes. For the purpose to define the sweet area for diffuse-field synthesis with discrete sources, the Chebyshev-type quadratures from \cite{graef,graefpotts,graefurl} were chosen for discretization. The parameter $t= 2N+1$ is chosen twice as high as $N$ plus one, consistently with the $t$ chosen for discretizing the circle before.

Simulation yields results in \autoref{fig:simu6} that are similar to those for the circle discretized by a $t=2N+1$ design in \autoref{fig:simu2}. Only now for the sphere, the number of sources needs to be much larger with $L=\{6,12,24,70\}$ for $N=\{1,2,3,5\}$ and $t=\{3,5,7,11\}$, when compared to $L=\{4,6,8,12\}$ for the circle. Hereby, the sweet-area radius estimated in eq.~\eqref{eq:sweet_x} also nicely models the $\psi=90\%$ diffuseness contour, as indicated by the orange red, dotted circle.
Concerning the sound energy density, the $\SI{1}{dB}$ contour stays outside this sweet area until \autoref{fig:simu6}(c) with a discretization by $24$ sources. 
The slightly non-constant sound pressure level measured by $\rho\,c^2\,w$ appears to be negligible in practical implementations using $N<4$ or $t$-designs with $t<9$.

\section{Diffuse sound fields of 2.5D WFS }
\label{sec:wfs}
Wave field synthesis (WFS)~\cite{berkhout,start,spors2013,ahrens,firtha} with a spherical/circular loudspeaker system of the radius $R_\mathrm{s}$ is able to reproduce primary, virtual sound sources at multiples of this radius $R_0=m\,R_\mathrm{s}$, below the spatial aliasing frequency for which the loudspeaker density of the practical implementation is higher than half the acoustic wavelength. One would expect improved diffuse field rendering from a virtual distance increase.
We assume a circular layout of such uncorrelated virtual sources to represent a diffuse sound field, using the global direction $\bm u_0$ from the origin to define their positions 
$\bm x_0=R_0\,\bm u_0$.
\begin{figure}[t]
    \centering
    \includegraphics[width=8.4cm]{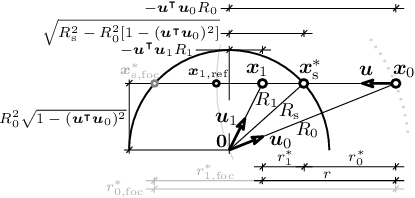}
    \caption{2.5D WFS scenario with uncorrelated virtual primary sources at the circular radius $R_0$, of which one is at $\bm x_0$ and  reproduced by secondary sources at the radius $R_\mathrm{s}$. Reproduction at $\bm x_1$ is dominated by the stationary-phase point $\bm x_\mathrm{s}^*$ and a reference contour through $\bm x_{1,\mathrm{ref}}$.}
    \label{fig:wfs-sketch}
\end{figure}

A similar aspect as discussed above is relevant here: When the space dimensions $D'$ occupied by the loudspeaker layout are fewer than the actual dimensions $D$ of the space, then primary sound sources are reproduced with erroneous decay $\frac{1}{\sqrt{r}^{D'-1}}$ that does not match
$\frac{1}{\sqrt{r}^{D-1}}$. 2D WFS would be the ideal choice in theory, requiring a horizontal circle of vertical line sources, but the affordable practical choice usually is a circle of point sources for 2.5D WFS, cf.~\cite{start}.  While a plane-wave sound field has a constant amplitude, its 2.5D WFS reproduction exhibits some amplitude decay, cf.~\cite{ahrens,sfstoolbox}. The introductory example \autoref{fig:simu1}(c) crudely approximated it by the amplitude decay of a horizontal line source.

More precisely, as sketched in \autoref{fig:wfs-sketch}, the part $r_1^*$ of the distance $r=\|\bm x_1-\bm x_0\|=r_0^*\pm r_1^*$ between observer at $\bm x_1=R_1\,\bm u_1$ and primary source at $\bm x_0=R_0\,\bm u_0$ often defines the 2.5D WFS local amplitude decay more than the distance $r$ to the primary source. In particular, $r_1^*$ is the distance $r_1^*=\|\bm x_\mathrm{s}^*-\bm x_1\|$ to the stationary-phase source that is located at $\bm x_\mathrm{s}^*$, or for a focused source $\bm x_\mathrm{s,foc}^*$, see~\cite{start,firtha}. This source dominates the amplitude received at $\bm x_1$ for the virtual source direction $\bm u_0$, and its amplitude can be corrected  for observation at a reference distance $r_{1,\mathrm{ref}}^*$ along the direction $\bm u$. Doing so for all pairs of stationary-phase points $\bm x_\mathrm{s}^*$ and associated reference points defines a reference contour of correct synthesis~\cite{firtha_referencing}. We choose a reference circle of the radius $R_0$ around the primary source, through the origin $\bm 0$, here. To lie on this contour, the distance $r$ must be $R_0$. 

\begin{figure}[b]
	\centering
	\includegraphics[height=40mm,trim=3mm 4mm 3mm 3mm, clip]{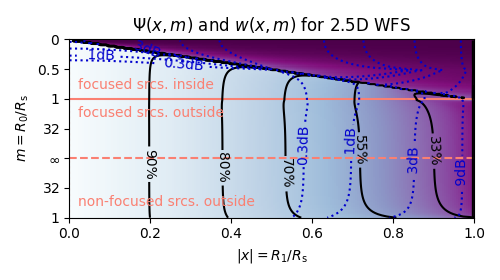}
	\includegraphics[height=37mm,trim=0mm -6mm 0mm 5mm,clip]{colorbar.png}
	\caption{Diffuseness $\psi$ (colors/solid contours) at off-center positions $x$ in 2.5D WFS of a circle of uncorrelated virtual sources (focused/non-focused) at different radii $R_0=m R_\mathrm{s}$ and sound energy density $w$ (dotted).
		\label{fig:wfs}}
\end{figure}

\autoref{apdx:wfs} outlines the underlying calculations and yields the stationary-phase approximated 2.5D WFS of a Green's function at $\bm x_0=R_0\,\bm u_0$ in eq.~\eqref{eq:g24dwfs}, 
\begin{align}
    G_\mathrm{2.5DWFS}=\sqrt{\frac{R_0- r_{0}^*}{r- r_0^*}}\sqrt{\frac{R_0}{r}}\;e^{\mp\mathrm{i}k\,r},\label{eq:gbwfs}
\end{align}
where $r_0^*$ yields with the constraint $\|\bm x_\mathrm{s}^*\|^2=R_\mathrm{s}^2$ and its coordinates $\bm x_\mathrm{s}^*=R_0\bm u_0\mp r_0^*\,\bm u$,
\begin{align}
    r_0^*&=-R_0\,\bm u_0^\intercal\bm u\mp \sqrt{R_\mathrm{s}^2-R_0^2\,[1-(\bm u_0^\intercal\bm u)^2]} .\label{eq:r1}
\end{align}
We apply the integrals of eqs.~\eqref{eq:I_calG} and \eqref{eq:w_calG} to superimpose such uncorrelated 2.5D WFS sources.
The resulting expressions are too elaborate for analytic integration, so our investigation relies on numerical integration over the angular integration variable $\mathrm{d}\Omega_0(\bm u_0)$, where $\bm u_0$ defines the primary sources $\bm x_0=R_0\bm u_0$, $R_\mathrm{s}=1$, $R_0=m\,R_\mathrm{s}$, $\bm u=\frac{\bm x_0-\bm x_1}{r}$, and eq.~\eqref{eq:r1} defines the magnitude square of eq.~\eqref{eq:gbwfs}, as used in $\mathcal{G}\propto|G_\mathrm{2.5DWFS}|^2$, using the integrals derived for $\rho\,c^2\,w$ eq.~\eqref{eq:w} and $\rho\,c\,\bm I$ eq.~\eqref{eq:I}, so that
\begin{align}
   \rho\,c^2\,w&=\frac{1}{L}\sum_{l=1}^L
     |G_{\mathrm{2.5DWFS},l}|^2,\\
   \rho\,c\,\bm I&=\frac{1}{L}\sum_{l=1}^L
     \bm u_l\,|G_{\mathrm{2.5DWFS},l}|^2,
\end{align}
and hereby $\psi=1-\frac{\rho\,c\,\|\bm I\|}{\rho\,c^2\,w}$.
\autoref{fig:wfs} displays the resulting sound energy density (dotted contour) and diffuseness (color/solid contour) for varying relative size $m$ of an uncorrelated virtual source circle, either non-focused or focused. Non-focused sources are equivalent to uncorrelated point-source loudspeakers without WFS when $m=1$, and to the typical 2.5D WFS of virtual plane waves with $m\rightarrow\infty$, cf.~\autoref{apdx:wfs}.

It has been a common belief that 2.5D WFS with plane-wave sound objects has benefits in producing diffuse sound fields. Listening experiments by Melchior et al.\ \cite{Melchior2008} and the crude approximation  \autoref{fig:simu1}(c) provide reason for doubt that the map in \autoref{fig:wfs} now clearly confirms: focused sources with $m<1$ synthesize a non-diffuse zone $|x|>m$ and can be disregarded, at small displacements $|x|\leq0.2$, diffuseness is still $\psi>90\%$ for any $m\geq1$; elsewhere, any $m>1$ even reduces diffuseness as its $70\%$, $55\%$,  $33\%$ contours in \autoref{fig:wfs} shrink. Conversely, sound pressure levels get more constant by the expanded $\SI{0.1}{dB}$, $\SI{1}{dB}$, $\SI{3}{dB}$ contours for $m\rightarrow\infty$ of non-focused sources, and even more so for $m\rightarrow1$ with focused sources.

\begin{figure*}[]
	\centering
	\includegraphics[height=5cm]{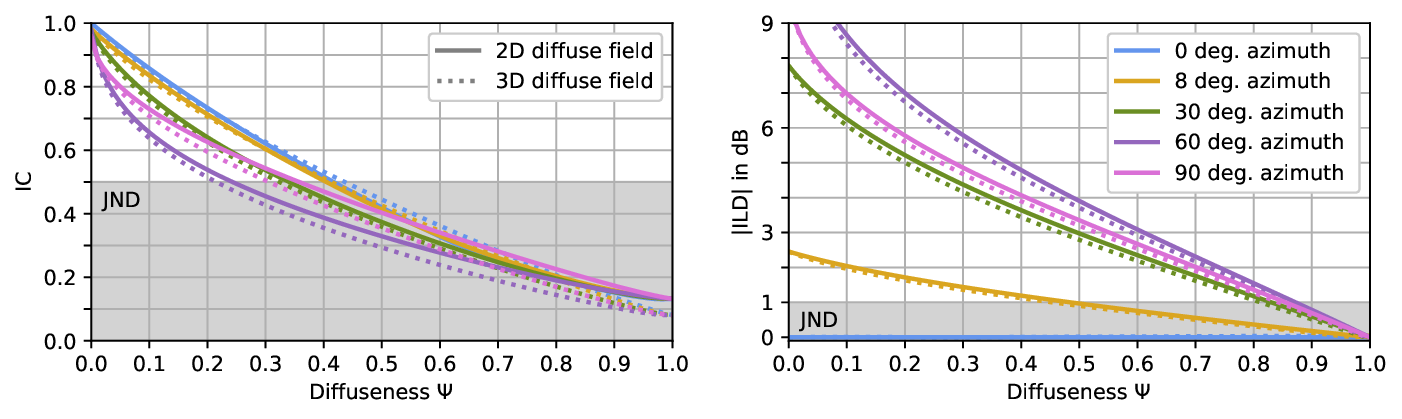}
	\caption{Interaural coherence (IC) and interaural level difference (ILD) evaluated for sound fields with variable diffuseness $\psi$. The evaluation uses a variably-weighted superposition of a 2D/3D diffuse sound field and a single-DOA soundfield from a horizontal, fromtal azimuth angle. 
		Gray areas indicate the JND to a diffuse-field reference. \label{fig:IC_ILD_Diffuseness}}
\end{figure*}

At high frequencies, sound field synthesis exhibits spatial aliasing that makes the observed loudspeaker phases appear random, so that the metrics for $m=1$ and non-focused sources are expected to apply.  

In summary, the effort of driving a ring of point-source loudspeakers via 2.5D WFS to reproduce uncorrelated virtual sources at other radii is surprisingly unsuccessful and even deteriorates diffuseness. While wave fronts of the virtual sources are successfully reproduced, their $\nicefrac{1}{r}$ or $\nicefrac{1}{\sqrt{r}}$ magnitude reproduces as $\nicefrac{1}{(\sqrt{r}\sqrt{r-r_0^*})}$, instead. Loudspeakers should be vertical line sources to enable 2D~WFS that accurately reproduces the target decay $\nicefrac{1}{\sqrt{r}}$.

\vspace*{-3mm}
\section{Perceptual Aspects}
\label{sec:discussion}
\vspace{-1mm}

The numerical simulations and analytical derivations could show which constellations favorably achieve a maximum of diffuseness across the entire audience area. Perceptually, this is equivalent to maintaining minimum the interaural level difference (ILD) and interaural coherence (IC) across this area. This section demonstrates the relation between diffuseness in the range $0\%\leq\psi\leq100\%$ and the resulting ILD/IC values and establishes the relation to known perceptual thresholds \cite{riedel2022surrounding, hartmann2002interaural, pollack1959binaural}. To this end, the ear signals of a uniformly diffuse sound field is simulated and superimposed with a unidirectional sound field (single non-diffuse direction from horizontal plane and frontal hemisphere), as numerically analyzed in~\autoref{fig:IC_ILD_Diffuseness}. The interaural cues are computed from Neumann KU100 HRTFs \cite{bernschutz2013spherical}, averaging across gammatone frequency bands of one equivalent rectangular bandwidth (ERB) between 200 Hz and 12.8 kHz. While the low-end JND $0.5$ of IC is uncritical in most cases and not surpassed for diffuseness of values of $\psi > 40\%$ for all directions, the ILD JND of around $\SI{1}{dB}$ is the more critical cue. A perceptually irrelevant ILD requires a diffuseness as high as $\psi > 80\%$ when the unidirectional component is lateral. No ILD is caused by unidirectional content from the median plane, where diffuseness values $\psi>40\%$ are still sufficiently high for low IC.

It is a lucky practical coincidence that while the vertical intensity component of a hemispherical loudspeaker layout drastically reduces the diffuseness to $\psi\approx50\%$, cf.\ \autoref{fig:hemisphere}, it will still be providing an acceptably small IC, and ILD will stay limited as long as intensity is not inclined by more than $8^\circ$ with regard to the median plane, either because of a listener who rolls the head sideways, or by a slightly inclined intensity, cf.~\autoref{fig:IC_ILD_Diffuseness}.

The most suitable solutions providing $100\%$ diffuseness rendering (circle/sphere of $\nicefrac{1}{\sqrt{r}^{D-1}}$) seem to be incapable of supplying perfectly constant sound pressure level everywhere inside. However, it seems from the respective simulations above that restricting the playback area to about $r<80\%\,R$ in size will keep the level constant enough, with variations below $\SI{1}{dB}$, which resembles relevant loudness JNDs~\cite{Zwislocki86,Slade23} or lies below.

Unless working with 2D WFS and plane-wave virtual sources, isotropy will typically also not be reached everywhere, but it is at least apparent that the more spherical and full-dimensional the layout becomes, and the larger in loudspeaker count, the easier it is to maintain isotropy around the center location, see also~\cite{riedel2022surrounding}.








\vspace*{-3mm}
\section{Conclusion}
\vspace{-2mm}
In this article we discussed theoretical requirements and limitations of diffuse sound fields synthesized with uncorrelated sources in the free field. The findings are of practical relevance when attempting to produce extended diffuse fields with loudspeakers, either in the anechoic chamber or in typical spatial audio rendering scenarios with loudspeakers or virtual sources.

We could prove there being optimal layouts that can synthesize vanishing sound intensity in the entire field enclosed, if their surface normal is rotation/shift invariant. This was accomplished by statistical expectation, assuming a continuous layer of uncorrelated surrounding sources of equal variance, relating sound intensity synthesized by the Green's function of the Helmholtz equation to Gau{\ss}' divergence theorem and Newton's spherical shell theorem for the Green's function of the potential equation. These are optimal layouts:

\begin{itemize}
    \item 
A uniform spherical layer (3D) of uncorrelated point sources is able to produce vanishing sound intensity, i.e.\ $100\%$ diffuseness, everywhere inside.

\item
A circular layout of uncorrelated (vertical line) sources produces $100\%$ diffuseness everywhere inside, in 2D. Embedded in 3D, it must be of infinite extent in the axis perpendicular to the circle, forming a cylindrical source distribution. Sources need to be uncorrelated along the circle, not necessarily along the cylinder axis. 

\item
A pair of opposing (plane) sources produces $100\%$ diffuseness everywhere inside in 1D. Embedded in 3D, they are two uncorrelated infinite-plane source distributions, whose two sides need to be uncorrelated, not necessarily the sources along either plane.
\end{itemize}

For these optimal source layouts, we could prove that energy density or sound pressure level will only be constant inside the 1D layout of parallel planes. A distance decay modified from the physical one $\nicefrac{1}{\sqrt{r}^{D-1}}$ to $\nicefrac{1}{r^\beta}$ was employed in the source integrals and solved by hypergeometric functions to show  that $\nicefrac{1}{\sqrt{r}^{D-1}}$ of physical sources yields $100\%$ diffuseness, while $\nicefrac{1}{\sqrt{r}^{D-2}}$ or $\nicefrac{1}{r^0}$ yields  constant sound pressure level. Our formally strict proof exploited $\nicefrac{1}{r^\beta}$ as generating function of the orthogonal Gegenbauer polynomials, and in particular the orthogonality between the constant potential of the zeroth-order polynomial to higher-order polynomials, given the integration weight $\varpi_\beta$ of the $D$-dimensional geometry.

We could moreover prove that it is impossible to construct a diffuse sound field that is isotropic regardless of the observer position within a set of uncorrelated sources at finite distance, with fixed source amplitude density $\sigma$. Isotropy requires a constant ratio $\sigma^2/\cos\phi$ of squared source amplitude density $\sigma^2$ and direction cosine $\cos\phi$ between the sound ray observed and the surface normal at location of its source. While $\sigma^2$ can be chosen to provide isotropy at a specific observer position, the direction cosines $\cos\phi$ are always position-dependent.

Non-circular/non-spherical source layouts exhibit a varying distance $R$ to the center of the layout. For elliptic 2D or ellipsoidal 3D layouts of directionally uniform source density, we could show that driving their uncorrelated sources by a gain $\sigma=\sqrt{R}^D$ produces $100\%$ diffuseness everywhere within. This choice differs from $\sigma=\sqrt{R}^{D-1}$ that provides isotropy to a central observer. 

For more arbitrary source layouts, we proposed a mode matching approach to find $\sigma$ on the example of a rectangular/cuboid superellipse to produce $100\%$ diffuseness inside, based on circular/spherical harmonics. 
The mode-matching solutions of these cases differ from $\sqrt{R}^{D}$, even more so from $\sigma=\sqrt{R}^{D-1}$, which still provides isotropy to a central observer. 

Hemispherical layouts are practically relevant in 3D audio and were shown to only reach $50\%$ diffuseness, due to the intensity component leaving the hemisphere. However, there will not be a tangential intensity component along the terminating plane of the hemisphere. 

For optimal distance decay $\nicefrac{1}{\sqrt{r}^{D-1}}$, we could show that circular/spherical $t$-designs as discrete source layouts yield sweet areas of $>90\%$ diffuseness within $r<\frac{N}{N+1}$, when choosing $t=2N+1$ for both cylindric and spherical layouts, using Gegenbauer polynomials.

Scaling the radius of the source layout while using only a limited central area is obviously always a stabilizing option. This may succeed virtually by array processing or soundfield synthesis.
It fails in the particular practical example of 2.5D WFS and a physical point-source loudspeaker layout: Uncorrelated virtual plane waves approximated by virtual sources at the radius 
$R_0\rightarrow\infty$ render limited diffuseness that is not better than what uncorrelated physical sources $R_0\rightarrow R_\mathrm{s}$ at the loudspeaker distance accomplish. Simulations of various scalings $R_0=m\,R_\mathrm{s}$ indicated that the 2.5D WFS technique has no benefits. Alternatively, theory suggests horizontal-only 2D~WFS to work. It employs vertical line-source  loudspeakers. While these are also optimal for any other 2D spatial audio rendering method, two extra benefits are accessible via 2D WFS when driven by uncorrelated virtual plane-wave sources: (i) a flat sound pressure level and (ii)~2D isotropy everywhere inside, cf.\ e.g.~\cite{tanaka23,ZotterSchultzGoellesWFSgit}.

Finally, the relation to relevant interaural measures as spatial auditory cues and their JNDs was established.  A simulation example showed that high diffuseness $\psi$ enforces low interaural coherence (IC) and level difference (ILD), both important for diffuse envelopment. It may therefore be beneficial to work with $\nicefrac{1}{\sqrt{r}^{D-1}}$ sources for optimal diffuseness and accept the synthesis of an imperfect but reasonably flat sound pressure level varying by only $1$ or $\SI{2}{dB}$. A strict diffuseness criterion of $\psi\geq80\%$ or $90\%$ is reasonable under the presence of a lateral sounds, which easily trigger noticeable ILDs. By contrast, vertical and hereby non-interaural directional content produced by upper hemispherical layouts still appears acceptable, despite imposing an upper limit $\psi\leq50\%$ to diffuseness.

\vspace*{-1mm}
\section{Acknowledgments}\vspace{-2mm}
We gratefully received funding from the Austrian Science Fund (FWF): P 35254-N (Envelopment in Immersive Sound Reinforcement, EnImSo).

\vspace*{-1mm}
\section{Data Availability Statement}\vspace{-2mm}
The jupyter lab python source code to create all sound field graphs shown in this article is available online~\cite{ZotterGit}, and a brief paper on the WFS basics that helped to develop the equations in \autoref{sec:wfs} is available online~\cite{ZotterSchultzGoellesWFSgit}.




\vspace{-3mm}
\bibliographystyle{IEEEtran}
\bibliography{refs.bib}

\begin{appendix}
\section{High-frequency Green's function}\label{apdx:hf_green}
Knowing that the free-field Green's function only depends on $r=\sqrt{\sum_{i=1}^Dx_i}$, we may define it via the Laplacian in terms of the radius $r$, only
\begin{align}
     \left[
     \frac{\partial^2}{\partial r^2}+\frac{D-1}{r}\frac{\partial}{\partial r} 
    +k^2
    \right]G(r)=-\delta(\bm x).
\end{align}
The corresponding homogeneous equation simplifies by multiplication with $r^2$ and substition $x=kr$ to 
\begin{align}
    \left[x^2\frac{\mathrm{d}^2}{\mathrm{d}x^2}+
   (D-1)x\frac{\mathrm{d}}{\mathrm{d}x}+x^2\right]y=0.
\end{align}
According to Sommerfeld~\cite[Ch.V(A.IV,C)]{Sommerfeld_engl}, it is useful to transform it into a Bessel differential equation
$\left[x^2\frac{\mathrm{d}^2}{\mathrm{d}x^2}+
   x\frac{\mathrm{d}}{\mathrm{d}x}+(x^2-\nu^2)\right]v=0$,
which succeeds by a suitable choice of $\nu$ in
$y=x^{-\nu} v$,
\begin{align}
   \bigg[x^2\frac{\mathrm{d}^2}{\mathrm{d}x^2}+
   (-2\nu+D-1)x\frac{\mathrm{d}}{\mathrm{d}x}+x^2\nonumber\\
  \qquad\quad -\nu(-\nu-1+D-1)\bigg]v=0,\nonumber
\end{align}
requiring the factor of $x\frac{\mathrm{d}}{\mathrm{d}x}$ to become unity $D=2\nu+2$
\begin{align}
   \left[x^2\frac{\mathrm{d}^2}{\mathrm{d}x^2}+
   x\frac{\mathrm{d}}{\mathrm{d}x}+x^2-\nu^2\right]v=0,\nonumber
\end{align}
so $\nu=\frac{D}{2}-1$,
making a suitably scaled and second-kind Hankel function~\cite[10.2.E6]{dlmf} a solution
\begin{align}
    G=a\,x^{-\nu}\,H_{\nu}^{(2)}(x)
\end{align}
that fulfills Sommerfeld's radiation condition~\cite[§28]{Sommerfeld_engl}.
The scalar $a$ is found by letting $k\rightarrow0$ and integrating eq.~\eqref{eq:green} over a small volume, with $\mathcal{S}_{D-1}=\frac{2\,\pi^{\frac{D}{2}}}{\Gamma(\frac{D}{2})}$,
\begin{align}
\lim_{k,V\rightarrow0}\int_V(\bigtriangleup+k^2)G\,\mathrm{d}V=-\int\delta\,\mathrm{d}V&=-1\nonumber\\
\lim_{k,V\rightarrow0}\int_V\bigtriangleup G\,\mathrm{d}V=
\lim_{k,r\rightarrow0} r^{D-1}\mathcal{S}_{D-1} \frac{\partial G}{\partial r}&=-1,\nonumber\\
\lim_{k,x\rightarrow0} r^{D-1}\mathcal{S}_{D-1}\frac{k\,\partial G}{\partial x}&=-1\nonumber\\
\lim_{k,x\rightarrow0} xr^{D-2}\mathcal{S}_{D-1}\frac{\partial G}{\partial x}&=-1\nonumber\\
\lim_{k,x\rightarrow0} xr^{2\nu}\frac{2\pi^{\nu+1}}{\Gamma(\nu+1)}\frac{\partial G}{\partial x}&=-1\nonumber.
\end{align}
The small-argument approximation of the Hankel function in $G=\frac{a}{x^\nu}H_\nu^{(2)}(x)$ yields, cf.~\cite[10.7.E7]{dlmf}, 
\begin{align}
    \lim_{x\rightarrow0}H_\nu^{(2)}(x)
    =\frac{\mathrm{i}\Gamma(\nu)}{\pi}\frac{2^\nu}{x^\nu}\nonumber,\\
    \lim_{x\rightarrow0}G
    =a\frac{\mathrm{i}\Gamma(\nu)}{\pi}\frac{2^\nu}{x^{2\nu}}\nonumber,\\
\lim_{x\rightarrow0}\frac{\partial G}{\partial x}
    =-2\mathrm{i} a\frac{\nu\Gamma(\nu)}{\pi}\frac{2^\nu}{x^{2\nu+1}},\nonumber
\end{align}
so that its scaling is found to be 
\begin{align}
xr^{2\nu}\frac{2\pi^{\nu+1}}{\Gamma(\nu+1)}2\mathrm{i} a\frac{\nu\Gamma(\nu)}{\pi}\frac{2^\nu}{x^{2\nu+1}}=1\nonumber\\
(2\pi)^{\nu}4\mathrm{i} a\frac{r^{2\nu}}{x^{2\nu}}=1\nonumber
\\
a=-\frac{\mathrm{i}}{4}\left(\frac{k^2}{2\pi}\right)^\nu.\nonumber
\end{align}
This yields together with $x^{-\nu}$
\begin{align}
    G&=-\frac{\mathrm{i}}{4}\left(\frac{k}{2\pi\,r}\right)^{\frac{D}{2}-1}\,H_{\frac{D}{2}-1}^{(2)}(kr),
\end{align}
and the far-field approximation~\cite[10.2.E6]{dlmf} 
\begin{align}
    \lim_{r\rightarrow\infty}
    H_\nu^{(2)}(z)&=\sqrt{\frac{2}{\pi\,z}}e^{-\mathrm{i}(z-\frac{(2\nu+1)\pi}{4})}\nonumber
\end{align}
yields the far-field Green's function eq.~\eqref{eq:green_hf} in \autoref{sec:source_distribution} 
\begin{align}
    \lim_{r\rightarrow\infty}G&=\frac{1}{4}\left(\frac{k}{2\pi\,r}\right)^{\frac{D}{2}-1}\,\sqrt{\frac{2}{\pi\,kr}}e^{-\mathrm{i}kr+\mathrm{i}\frac{(D-3)\pi}{4}}\nonumber\\
    &=
    \sqrt{\frac{2\pi r}{\mathrm{i}k}}^{\frac{3-D}{2}}\,\frac{e^{-\mathrm{i}kr}}{4\pi\,r}.\label{eq:hfgreen}
\end{align}

\section{Reduced decay for infinite extent}
\label{apdx:excess}
The reduced distance decay distributed sources exhibit in terms of a parallel distance $r$ with regard to $E$ infinitely extended excess dimensions is found by first regarding a negative power of $r$
\begin{align}
    r^{-2\alpha}=\left[\sum_{j=1}^{D'}(x_i-x_{\mathrm{s},i})^2+\sum_{i=D'+1}^D x_{i}^2\right]^{-\alpha}.
\end{align} 
We differentiate $r$ it in all the $E$ excess dimensions
\begin{align}
    \left[\prod_{i=1}^E\frac{\partial}{\partial x_{D'+i}}\right]\frac{1}{r^{2\alpha}}
    =
    \frac{(-2)^{E}(\alpha+E-1)!}{(\alpha-1)!}\frac{\prod_{i=1}^Ex_{D'+i}}{r^{2\alpha+E}}\nonumber.
\end{align}
To make the right hand side $\frac{1}{r^{2\alpha+E}}$ match $\mathcal{G}'_D\propto \frac{1}{r^{D-1}}$, its exponent  needs to be $2\alpha+E=D-1$.
Because of the $E$ derivatives, integrating over the $E$ excess dimensions is trivial, just removes the differentials, and yields $\frac{1}{r^{2\alpha}}$ on the left, so that the decay exponent correspondingly reduces from $2\alpha+E=D-1$ to $2\alpha=D-E-1=D'-1$.
This relation is used in \autoref{sec:subspace} eq.~\eqref{eq:reduced_decay_excess}.

\section{Modified decay exponent $\beta$}
\label{apdx:hypergeometric}
\autoref{sec:modified_beta} uses integrals to find $\rho\,c^2\,w$ and $\rho\,c\,I_\mathrm{x}$ for an  observer shifted along the first coordinate $x_1=x$ within a spherical shell of $D$ dimensions, involving the Euclidean distance  $r=\|\bm x-\bm x_\mathrm{s}\|=\sqrt{1-2\cos\phi\,x+x^2}$, where the direction cosine between source and observer location along the axis $x_1$ is summarized as $z=\cos\phi$.
Integration over the angular element $\frac{\mathrm{d}\Omega_0}{\mathcal{S}_{D-1}}=\varpi\,\mathrm{d}z$ with $-1\leq z\leq 1$ from the center of the sphere  uses the weight $\varpi=\frac{\mathcal{S}_{D-2}}{\mathcal{S}_{D-1}}\sqrt{1-z^2}^{D-3}$, cf.~\cite{HilbertCourant}. The axisymmetric integral around the first dimension of the shift $x$ remains over the decay $\frac{1}{{r}^{2\beta}}$ for sound energy density 
and additionally over the directional factor $\frac{z-x}{r}$ in
$\frac{z-x}{r^{2\beta+1}}$
for the intensity $\rho\,c\,I_\mathrm{x}$ along $x_1$,
\begin{align}
   \rho\,c^2\,w&=\frac{1}{\tilde N}\int_{-1}^1\frac{\sqrt{1-z^2}^{D-3}\mathrm{d}z}{\sqrt{1-2zx+x^2}^{2\beta}},\label{eq:wz}\\
     \rho\,c\,I_\mathrm{x}&=
     \frac{1}{\tilde N} \int_{-1}^1\frac{(z-x)\,\sqrt{1-z^2}^{D-3}\mathrm{d}z}{\sqrt{1-2xz+x^2}^{2\beta+1}}.\label{eq:Iz}
\end{align}
These two integrals 
are better solved analytically than numerically to obtain numerically stable results.
We substitute $z=2t-1$ and get 
\begin{align}
    \rho\,c^2\,w
    \propto&\frac{1}{(1+x)^{2\beta}}\int_{0}^1\frac{t^\frac{D-3}{2}(1-t)^\frac{D-3}{2}\mathrm{d}t}{[1-\frac{4x}{(1+x)^2}t]^\frac{2\beta}{2}},
    \\
   \rho\,c\,I_\mathrm{x}\propto&
   \frac{1}{(1+x)^{2\beta}}\int_{0}^1\frac{t^\frac{D-3}{2}(1-t)^\frac{D-3}{2}\mathrm{d}t}{[1-\frac{4x}{(1+x)^2}t]^\frac{2\beta+1}{2}}\\
   &
    -\frac{2}{(1+x)^{2\beta+1}}\int_{0}^1\frac{t^\frac{D-1}{2}(1-t)^\frac{D-3}{2}\mathrm{d}t}{[1-\frac{4x}{(1+x)^2}t]^\frac{2\beta+1}{2}},
   \nonumber
\end{align}
so that the integral related to $F(a,b;c;z)$ \cite[15.2.E1]{dlmf}
\begin{align}
    F(a,b;c;z)&=\frac{\Gamma(c)}{\Gamma(b)\Gamma(c-b)}\int_{0}^1\frac{t^{b-1}\,(1-t)^{c-b-1}}{(1-zt)^a}\;\mathrm{d}t,\nonumber
\end{align}
helps to find the expressions of eqs.~\eqref{eq:w_beta} and \eqref{eq:I_beta}, which used $a=\frac{2\beta}{2}$ or $a=\frac{2\beta+1}{2}$, $b=\frac{D-3}{3}+1$ or $b=\frac{D-1}{2}+1$, and $c=\frac{D-3}{2}+b+1$.
The known value $F(a,b;c;0)=1$ supporting the normalization of $\rho\,c^2\,w(0)=1$ without additional factors, and symmetry-induced  $\rho\,c\,I_\mathrm{x}(0)=0$ at $x=0$ were used to verify scaling. The above equations also yield eq.~\eqref{eq:psi_beta} for diffuseness $\psi$.

\section{2.5D WFS}\label{apdx:wfs}
\begin{figure*}[t]
	\centering
	\includegraphics[width=3.7cm,trim=7mm 7mm 5mm 7mm,clip]{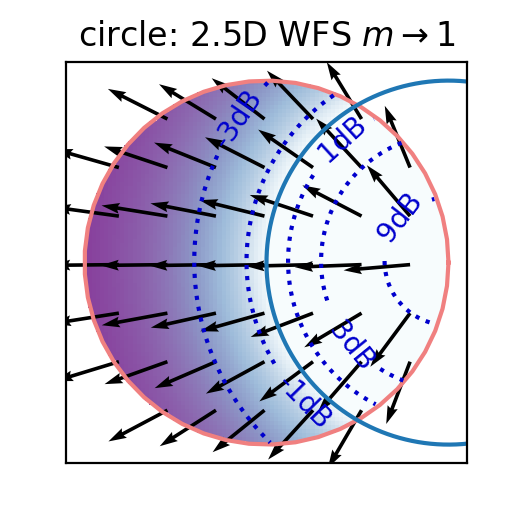}
	\includegraphics[width=3.7cm,trim=7mm 7mm 5mm 7mm,clip]{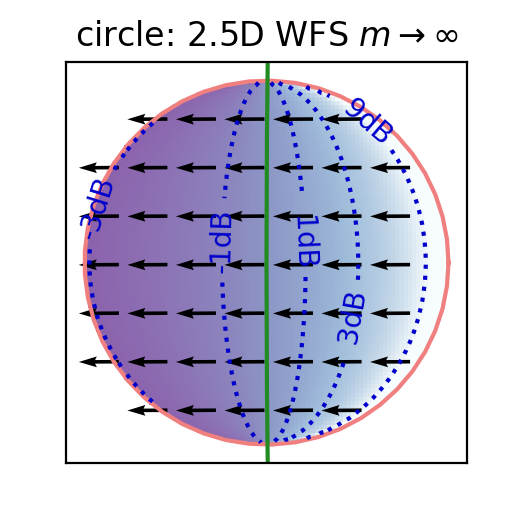}
	\includegraphics[width=3.7cm,trim=7mm 7mm 5mm 7mm,clip]{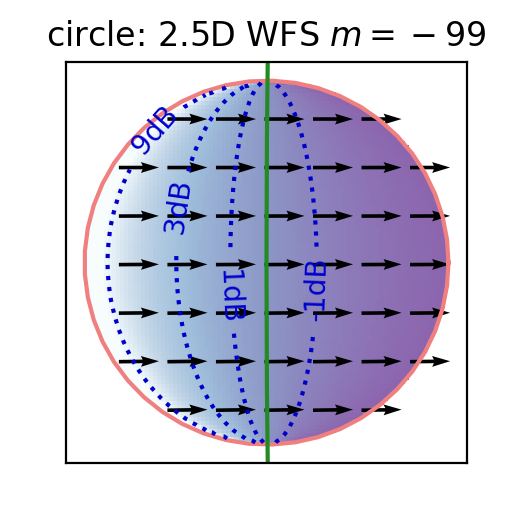}
	\includegraphics[width=3.7cm,trim=7mm 7mm 5mm 7mm,clip]{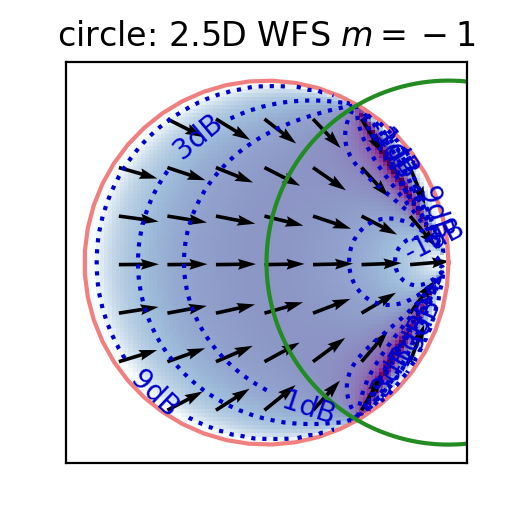}	
	\caption{Sound energy density (color/contours) and sound intensity (arrows) of stationary-phase approximated 2.5D WFS for non-focused (left $x_0=1,\infty$) and focused (right $x_0=\infty,1$) virtual source, $\SI{0}{dB}$ reference contour is green.\label{fig:25dwfs_src}}
\end{figure*}
Formal background on wave field synthesis (WFS) as analyzed here was prepared in~\cite{ZotterSchultzGoellesWFSgit}. WFS employs the Kirchhoff-Helmholtz integral, where $r_1=\|\bm x_\mathrm{s}-\bm x_1\|$ and $r_0=\|\bm x_\mathrm{s}-\bm x_0\|$ are: the distance $r_1$ between a secondary source $\bm x_\mathrm{s}$ of the convex, closed, linear contour $S$ and the observation position $\bm x_1$, and $r_0$ is the distance between the location $\bm x_\mathrm{s}$ of a secondary source 
$G(r_1)=\frac{e^{-\mathrm{i}kr_1}}{4\pi\,r_1}$
on the contour and the primary source $G(r_0)=\frac{e^{-\mathrm{i}k(\pm r_0)}}{4\pi\,r_0}$ located at $\bm x_0$, and evaluated at the contour. The primary source uses a negative phase in case of a focused sources with time-reversal, cf.~\cite{ahrens}. The synthesis integral 
\begin{align}
   p(\bm x_1)&=-\int_S \left[G(r_0)\nabla G(r_1)-G(r_1)\nabla G(r_0)\right]^\intercal\bm n\,\mathrm{d}S
\end{align}
should be $p=G(r)$ with $r=\|\bm x_0-\bm x_1\|$. The gradient $\nabla G(r_{0,1})=(\nabla r_{0,1})\;\frac{\mathrm{d} G(r_{0,1})}{\mathrm{d} r_{0,1}}$ becomes for $kr_0,kr_1\gg1$,
\begin{align}
\nabla G(r_{0,1})=\bm{\hat{u}}_{0,1}\,(\mp\mathrm{i}k-\,r_{0,1}^{-1})\,G(r_{0,1})
\end{align}
with regard to the contour coordinates $\bm x_\mathrm{s}$. With $\bm{\hat{u}}_{0,1}=\frac{\bm x_\mathrm{s}-\bm x_{0,1}}{r_{0,1}}$ 
and 
$\lim_{kr_{0,1}\rightarrow\infty}\nabla G(r_{0,1})=\mp\mathrm{i}k\,\bm{\hat{u}}_{0,1}\,G(r_{0,1})$,
\begin{align}
   p&\approx\mathrm{i}k\,\int_S \left[\bm{\hat{u}}_0\mp\bm {\hat{u}}_1\right]^\intercal\bm n\,G(\pm r_0)\,G(r_1)\, \mathrm{d}S.\label{eq:ffwfs}
\end{align}
Integration shall be approximated via stationary-phase method~\cite{Skudrzyk71} for a 2D horizontal linear contour, yielding 
\begin{align}
   p&\approx\mathrm{i}k\,\frac{2\bm u^\intercal\bm n}{\bm u^\intercal\bm n}\,\sqrt{\frac{2\pi}{\mathrm{i}k}}
    \sqrt{\frac{1}{\pm \frac{1}{r_0^*}+\frac{1}{r_1^*}}}
   \frac{e^{-\mathrm{i}k(\pm r_0^*+r_1^*)}}{(4\pi)^2\,(\pm r_0^*\,r_1^*)},\nonumber\\
   &\approx\sqrt{\frac{\pm\mathrm{i}k}{2\pi}}\,
   \sqrt{\frac{r_0^*\pm r_1^*}{r_0^*\,r_1^*}}
   \;G(\pm r_0^*+r_1^*),\label{eq:spa_wfs}
\end{align}
for the phase $e^{-\mathrm{i}k(\pm r_0+r_1)}$ of the product $G(r_0)\,G(r_1)$, the integration element $\mathrm{d}S=\mathrm{d}s$ and the stationary-phase condition $\frac{\mathrm{d}(\pm r_0+r_1)}{\mathrm{d}s}=0$ defines the stationary-phase point $\bm x_\mathrm{s}^*$ to which the distances are $r_0^*$, $r_1^*$ and sum up to $r=r_0^*\pm r_1^*$, $\bm u=\bm {\hat{u}}_0=\mp\bm {\hat{u}}_1$, see \autoref{fig:wfs-sketch}. At the stationary-phase point, the second-order derivative is $\frac{\mathrm{d}^2(\pm r_0^*+r_1^*)}{\mathrm{d}s^2}=[\pm\frac{1}{r_0^*}+\frac{1}{r_1^*}](\bm u^\intercal\bm n)^2$, cf.~\cite{start,firtha,grandjean2,ZotterSchultzGoellesWFSgit}.

Synthesis obviously delivers a scaled version of $G(r)$. Unity magnitude is enforced by a factor $H$ for every stationary-phase point $\bm x_\mathrm{s}^*$, at least at a reference distance $r|_\mathrm{ref}=R_0$ from the primary source~\cite{firtha_referencing}, describing a reference contour through the origin $\bm 0$, here. While the second partial distance $r_0^*|_\mathrm{ref}\equiv r_{0}^*$ does not depend on the reference distance, the first one  $r_1^*|_\mathrm{ref}=r_{1,\mathrm{ref}}^*$ does. Eq.~\eqref{eq:spa_wfs} is equalized to unity at the reference distance $r=R_0$ and $r_1^*=r_{1,\mathrm{ref}}^*$ by 
$H=\sqrt{\frac{2\pi}{\pm\mathrm{i}k}}\sqrt{\frac{r_0^*\,r_{1,\mathrm{ref}}^*}{R_0}}(4\pi\,R_0)$,
so 2.5D WFS based on eq.~\eqref{eq:ffwfs} becomes, cf.~\cite{ZotterSchultzGoellesWFSgit}, 
\begin{align}
   p&\approx\mathrm{i}k\,\int_S \left[\bm{\hat{u}}_0\pm\bm{\hat{u}}_1\right]^\intercal\bm n\,G(r_0)\,G(r_1)\,H\, \mathrm{d}S\nonumber\\
   &\approx\,
   \sqrt{\frac{R_0-r_{0}^*}{r-r_0^*}}
  \sqrt{\frac{R_0}{r}}\; e^{\mp\mathrm{i}kr}.
  \label{eq:gwfs2d-first}
\end{align}
\autoref{sec:wfs} superimposes such uncorrelated virtual-source fields across directions for diffuse field synthesis. We introduce a reproduced source term  $p\rightarrow G_\mathrm{2.5DWFS}$ for simplicity. The corresponding 2.5D WFS Green's function eq.~\eqref{eq:gwfs2d-first} becomes, see also \autoref{fig:25dwfs_src},
\begin{align}
   G_\mathrm{2.5DWFS}&\approx \sqrt{\frac{R_0-r_{0}^*}{r-r_0^*}}
   \sqrt{\frac{R_0}{r}}\;e^{\mp\mathrm{i}k\,r}.\label{eq:g24dwfs}
\end{align}
When the non-focused virtual-source radius approaches the secondary-source radius $R_0\rightarrow R_\mathrm{s}$, the outer partial distance vanishes $r_0^*\rightarrow0$, 
\begin{align}
   \lim_{R_0\rightarrow R_\mathrm{s}}G_\mathrm{2.5DWFS}&=\frac{R_\mathrm{s}}{r}\;e^{-\mathrm{i}k\,r}=4\pi\,R_\mathrm{s}\;G_\mathrm{3D}(r).\nonumber
\end{align}
When the virtual-source distance extends to infinity $R_0\rightarrow\infty$, then the ratio $\sqrt{\frac{R_0}{r}}\approx1$ because directions align $\bm u\rightarrow\,\bm u_0$ and the point-line distance $r=R_0\mp R_1\bm u_1^\top\bm u_0\approx R_0$. The ratio $\sqrt{\frac{R_0-r_{0}^*}{r-r_0^*}}=\sqrt{\frac{r_{1,\mathrm{ref}}^*}{r_1^*}}$ is approximated by $r_{1,\mathrm{ref}}^*=\sqrt{R_\mathrm{s}^2-R_1^2[1-(\bm u_1^\top\bm u_0)^2]}$ and $r_1^*=r_{1,\mathrm{ref}}^*\mp R_1\,\bm u_1^\top\bm u_0$, so that
\begin{align}
   \lim_{R_0\rightarrow \infty}G_\mathrm{2.5DWFS}&=
   \sqrt{\frac{1}{1\mp\frac{R_1\,\bm u_1^\top\bm u_0}{\sqrt{R_\mathrm{s}^2-R_1^2[1-(\bm u_1^\top\bm u_0)^2]}}}}\;e^{\mp\mathrm{i}k\,r}.\nonumber
\end{align}
At high frequencies $kr\gg 1$, the gradient of $G_\mathrm{2.5DWFS}$ is
\begin{align}
   \nabla G_\mathrm{2.5DWFS}&\approx\mathrm{i}k\,\bm u\,G_\mathrm{2.5DWFS},\nonumber
\end{align}
so that the known integrals for $\rho\,c^2\,w$ and $\rho\,c\,\bm I$ can be used; a sign change of the direction $\bm u$ restores consistency with definitions above.

\end{appendix}

\end{document}